

\input harvmac
\input epsf

\noblackbox
\pageno=0\nopagenumbers\tolerance=10000\hfuzz=5pt
\baselineskip=12pt plus2pt minus2pt
\line{\hfill\tt hep-ph/9512208}
\line{\hfill CERN-TH/95-323}
\line{\hfill Edinburgh 95/560}
\vskip 12pt
\centerline{\bf UNIVERSALITY AND SCALING}
\centerline{\bf IN PERTURBATIVE QCD AT SMALL $x$}
\vskip 24pt\centerline{Stefano~Forte$^{(a)}$\footnote*{On leave
{}~from INFN, Sezione di Torino, Turin, Italy (address after December 1,
1995).}
and Richard D.~Ball$^{(a, b)}$\footnote{\dag}{Supported in
part by a Royal Society University Research Fellowship.}
}
\vskip 12pt
\centerline{\it ${}^{(a)}$Theory Division, CERN,}
\centerline{\it CH-1211 Gen\`eve 23, Switzerland.}
\smallskip
\centerline{\it ${}^{(b)}$Department of Physics and Astronomy}
\centerline{\it University of Edinburgh, Edinburgh EH9 3JZ, Scotland}
\vskip 36pt
{\narrower\baselineskip 10pt
\centerline{\bf Abstract}
\medskip
We present a pedagogical review of the universal scaling properties
displayed by the structure function $F_2$ at small $x$ and large $Q^2$
as measured at HERA. We first describe the derivation of the
double asymptotic scaling of $F_2$ from the leading-order
Altarelli-Parisi equations of perturbative QCD. Universal next-to-leading
order corrections to scaling are also derived. We explain why the universal
scaling behaviour is spoiled when the initial distributions rise too steeply
by considering the nonsinglet distribution $F_2^p-F_2^n$ as an explicit
example. We then examine the stability of double scaling
to the inclusion of higher order singularities, explaining how
the perturbative expansion at small $x$ can be reorganized in such a way that
each order is given by the sum of a convergent series of contributions
which are of arbitrarily high order in the coupling.
The wave-like nature of perturbative evolution is then shown to persist
throughout almost all the small $x$ region, giving rise asymptotically to
double scaling for a wide class of boundary conditions.
\smallskip}
\vskip 16pt
\centerline{Invited lectures\footnote\ddag{Presented by S.F.} at the}
\centerline{\it XXXV Cracow School of Theoretical Physics}
\centerline{Zakopane, Poland, June 1995}
\medskip
\centerline{\it to be published in the proceedings}
\vfill
\line{CERN/95-323\hfill}
\line{November 1995\hfill}
\eject


\def\neath#1#2{\mathrel{\mathop{#1}\limits_{#2}}}
\def\lsim{\mathrel{\rlap{\lower4pt\hbox{\hskip1pt$\sim$}}
    \raise1pt\hbox{$<$}}}         
\def\as{\alpha_s}

\def\Re{\,\hbox{Re}\,}

\def\etal{{\it et al.}}
\def\rhs{right hand side}

\def\toinf#1{\mathrel{\mathop{\sim}\limits_{\scriptscriptstyle
{#1\rightarrow\infty }}}}
\def\frac#1#2{{{#1}\over {#2}}}
\def\half{\hbox{${1\over 2}$}}
\def\quarter{\hbox{${1\over 4}$}}
\def\smallfrac#1#2{\hbox{${{#1}\over {#2}}$}}

\def\GeV{{\rm GeV}}
\def\MS{\hbox{$\overline{\rm MS}$}}

\catcode`@=11 
\def\slash#1{\mathord{\mathpalette\c@ncel#1}}
 \def\c@ncel#1#2{\ooalign{$\hfil#1\mkern1mu/\hfil$\crcr$#1#2$}}
\def\lsim{\mathrel{\mathpalette\@versim<}}
\def\gsim{\mathrel{\mathpalette\@versim>}}
 \def\@versim#1#2{\lower0.2ex\vbox{\baselineskip\z@skip\lineskip\z@skip
       \lineskiplimit\z@\ialign{$\m@th#1\hfil##$\crcr#2\crcr\sim\crcr}}}
\catcode`@=12 

\def\qns{q}

\def\PR{{\it Phys.~Rev.~}}

\def\NP{{\it Nucl.~Phys.~}}
\def\NPBPS{{\it Nucl.~Phys.~B (Proc.~Suppl.)~}}
\def\PL{{\it Phys.~Lett.~}}
\def\PRep{{\it Phys.~Rep.~}}

\def\SJNP{{\it Sov.~Jour.~Nucl.~Phys.~}}

\def\ZP{{\it Zeit.~Phys.~}}

\def\vol#1{{\bf #1}}\def\vyp#1#2#3{\vol{#1} (#2) #3}


\footline={\hss\tenrm\folio\hss}

\newsec{Structure Functions at Small $x$}
The small $x$ limit of structure functions measured in deep-inelastic
scattering is the frontier of perturbative QCD both from the experimental
and the theoretical point of view. In deep-inelastic scattering
experiments~\ref\guirev{See G.~Altarelli, \PRep\vyp{81}{1981}{1} for a general
introduction and notation.},
one measures the total, fully inclusive cross section for scattering of a
virtual photon with virtuality $-Q^2$ over a nucleon. The center-of-mass
energy of the collision is given by the Mandelstam invariant
$s=Q^2{1-x\over x}$. In the large $Q^2$ limit the cross section
is parametrized by a single form factor $F_2(x,Q^2)$ which
is determined by the underlying partonic degrees of freedom,
because its moments
with respect to $x$ are directly related to the nucleon matrix elements
of quark and gluon operators. The scale dependence of these moments
is governed by renormalization group equations which
summarize the dynamical content of perturbative QCD.

The small $x$ limit at large $Q^2$
thus corresponds to probing the light-cone dynamics of the nucleon
(large $Q^2$) in the high energy limit (large $s\sim{Q^2/x}$). This limit
stretches perturbative QCD towards its nonperturbative frontier. On the
one hand, one might expect the high-energy limit of total cross sections
to be governed by unitarity, and in particular by  the $t$-channel
exchange of Regge trajectories~\ref\landrev{See P.~Landshoff,
{\tt hep-ph/9410250} and references therein.}, which do not admit a
simple perturbative
interpretation.  On the other hand, perturbation theory itself in this
limit is somewhat problematic, hinting to  its eventual breakdown. Indeed,
the anomalous dimensions which govern the perturbative scaling behaviour
grow without bounds, implying that Bjorken scaling
is shifted to larger and larger values of $Q^2$. This is a manifestation
of the fact that there is now another large scale in the theory besides $Q^2$,
namely $s$ itself: hence the renormalization group will have to be
adapted in order to sum up this scale too. It is not {\it a priori} obvious
in which kinematic region this might be possible, if at all.

Experimentally, accessing this region requires the very high center-of-mass
energies that have only been attained very recently at the electron-proton
collider HERA\nref\dataa{ZEUS~Collab.,
\PL\vyp{B316}{1993}{412}\semi H1~Collab.,
\NP\vyp{B407}{1993}{515}.}
\nref\datac{ZEUS Collaboration, \ZP\vyp{C65}{1995}{379}\semi
H1 Collaboration, \NP\vyp{B439}{1995}{471}.}
\nref\datad{F.~Eisele, Summary talk at the
Europhysics Conference, Brussels, August 1995}~\refs{\dataa-\datad}.
When the first data on $F_2$ at large $Q^2$ and small $x$
were first presented they have provoked a considerable amount of surprise:
they not only displayed  the expected large violations of
Bjorken scaling, growing larger as $x$ decreases, rather,
they also deviated from the Regge behaviour which
is well tested by high-energy elastic scattering data, by displaying a
marked rise as $x$ decreases at fixed $Q^2$, whereas Regge theory
would have a flat or almost flat behaviour. Yet, such a violation of
Regge behaviour was predicted more than twenty years ago as a direct
consequence
of the leading-order renormalization group equations of perturbative
QCD~\ref\DGPTWZ{A.~De~R\'ujula et al., \PR\vyp{D10}{1974}{1649}}.

This non-Regge rise takes the form of a simple universal scaling
law~\ref\DAS{R.~D.~Ball and S.~Forte, \PL\vyp{B335}{1994}{77}.}
satisfied by $F_2(x,Q^2)$ at large $Q^2$ and small $x$: the structure
function depends only on a variable $\sigma(x,Q^2)$. Furthermore,
this dependence is universal, unlike Bjorken scaling, where asymptotically
structure functions depend only on $x$, but in a non-universal, uncalculable
way: hence it actually corresponds to a double scaling law, as the
universal dependence may be scaled out.
The way this double asymptotic scaling behaviour has arisen out of the HERA
data is shown in fig.~1, which displays $F_2$ as a function of the
scaling variable $\sigma(x,Q^2)$ (for all accessible values
of $x$ and $Q^2$), for each successive published set of data.
\nref\test{R.~D.~Ball and S.~Forte, \PL\vyp{B336}{1994}{77}.}
\nref\datab{G.~Wolf, for the ZEUS Collab.,
talk at the International Workshop on Deep
Inelastic Scattering. Eilat, Israel, February 1994\semi
K.~M\"uller, for the H1 Collab.,
talk at the 29th Rencontre de Moriond, March 1994.}\nfig\scaldat{Scaling
plot of the experimental data on the proton structure function $F_2(x,Q^2)$
{}~from (a) the 1992 HERA run~\dataa\ (adapted from ref.~\DAS) ,
(b)  the preliminary analysis of the 1993
HERA run~\datab\ (adapted from ref.~\test), (c) the 1993
run~\datac, (d) the preliminary analysis
of the 1994 run (H1 only) (adapted from ref.~\datad).
The diamonds are ZEUS data and the squares are H1 data.
Only points with $\rho,\,\sigma>1$ are included in the plots b)-d),
and in (d) only those with $Q^2>5$~GeV$^2$ (see text).
The straight line shown is the predicted asymptotic double
scaling behaviour (with fitted normalization).}

It is apparent that double asymptotic scaling is the foremost feature
of $F_2$ at small $x$ and large $Q^2$, at least in the region presently
explored by the HERA experiments. Understanding perturbative QCD in
this region thus means understanding the physics of double
asymptotic scaling. This entails understanding why the simple scaling
prediction, which, after all, follows from a leading-order renormalization
group analysis, survives the problems of perturbative
instability and need for the inclusion of other large scales alluded to
above.

In these lectures we will briefly summarize the current status of the
current theoretical and phenomenological understanding of double scaling.
In sect.~2 we will derive the double scaling prediction from
leading-order perturbative QCD and discuss the physics behind this
behaviour. We will then see how double scaling may be spoiled
if parton distributions at low scale are
too steep, taking as an example the case of nonsinglet
structure functions, where this actually happens. We will further explain how
double scaling is modified (but its universality preserved) by the inclusion
of next-to-leading order corrections and show that present-day data
are perfectly described by such a next-to-leading order analysis.
In sect.~3 we will then discuss the behaviour of the perturbative expansion
of anomalous dimensions when the effects of the other large scale which
is present in the problem are included to all orders in the coupling.
We will show that the perturbative expansion may be reorganized so
as to sum up these effects, and that double scaling emerges then
as the generic universal asymptotic behaviour. We will conclude by recalling
some recent phenomenological applications of this formalism to precision tests
of QCD, and summarizing the most promising future theoretical
and phenomenological developments.
\bigskip
\newsec{Double Asymptotic Scaling at Leading and Next-to-Leading Order}
\medskip
The perturbative evolution of the structure function $F_2(x,Q^2)$
is determined by first decomposing it into parton distributions:
\eqn\ftwodef{x\inv F_2(x,Q^2)\equiv\sum_{i=1}^{n_f} e^2_iC_i\otimes
\left(q_i+\bar q_i\right)+ C_g\otimes g,}
where $n_f$ is the number of active flavors, $e_i$ is the electric charge
of the quark distribution $q_i(x;Q^2)$, $\otimes$ denotes the convolution
with respect to $x$, i.e. $[f\otimes g](x)\equiv\int_x^1\!\smallfrac{dy}{y}\,
f\left(\smallfrac{x}{y}\right) g(y)$, and
the  coefficient functions at leading order are simply
\eqn\cf{C_i(x,Q^2)=\delta(1-x);\quad C_g(x,Q^2)=0,}
while at higher orders they depend on the specific choice of renormalization
prescription (factorization scheme): in most of the subsequent
treatment we will choose a scheme (parton scheme) in which eq.~\cf\
remains true to all orders. The evolution of parton distributions is
in turn determined by the Altarelli-Parisi equations~\ref\ap{
G.~Altarelli and G.~Parisi, \NP\vyp{B126}{1977}{298}.}
\eqn\aps{\eqalign{
\frac{d}{dt}\pmatrix{g\cr q_{\rm S}\cr}&=
\frac{\as(t)}{2\pi}
\pmatrix{P_{gg}&P_{gq}\cr P_{qg}&P_{qq}^{\rm S}\cr}
\otimes \pmatrix{g\cr q_{\rm S}\cr},\cr
\frac{d}{dt} q_{\rm NS}
&=\frac{\as(t)}{2\pi} P_{qq}^{\rm NS}\otimes q_{\rm NS},\cr}}
where $t\equiv\ln(Q^2/\Lambda^2)$ and the singlet and nonsinglet
quark distributions are respectively given by
\eqn\qpd{\eqalign{
q_{\rm S}(x,t)&
=\sum_{i=1}^{n_f}\left(q_i(x;Q^2)+\bar{q}_i(x;Q^2)\right),\cr
q_{\rm NS}(x,t)&=\sum_{i=1}^{n_f}
\left(\smallfrac{e_i^2}{\langle e^2\rangle}-1\right)
\big(q_i(x;Q^2)+\bar q_i(x;Q^2)\big),\cr}}
with $\langle e^2\rangle=\smallfrac{1}{n_f}\sum_{i=1}^n e^2_i$.
In terms of these distributions the decomposition \ftwodef\ becomes,
in a parton scheme~\cf, simply
\eqn\ftsx{F_2(x,Q^2)= \langle e^2\rangle
x\big(q_{\rm S}(x,t)+q_{\rm NS}(x,t)\big).}
The splitting functions depend on $t$ through  the strong coupling
$\alpha_s(t)$, and at $n$-th perturbative order are given by
\eqn\splf{{\alpha_s(t)\over2\pi}P(x,t)=\sum_{i=1}^n P^{(i)}
\left({\alpha_s(t)\over 2\pi}\right)^i+O\left(\alpha_s^{n+1}\right).}

The Altarelli-Parisi equations eq.~\aps\
are simply the inverse Mellin transforms of
the renormalization group equations satisfied by the nucleon matrix elements
of quark and gluon operators, namely
\eqn\rgeq{\eqalign{
\frac{d}{dt}\pmatrix{g(N,t)\cr q_{\rm S}(N,t)\cr}&=\frac{\as(t)}{2\pi}
\pmatrix{\gamma_{gg}\big(N,\as(t)\big)&
         \gamma_{gq}\big(N,\as(t)\big)\cr
         \gamma_{qg}\big(N,\as(t)\big)&
         \gamma_{qq}^{\rm S}\big(N,\as(t)\big)\cr}
\pmatrix{g(N,t)\cr q_{\rm S}(N,t)\cr},\cr
\frac{d}{dt}q_{\rm NS}(N,t)
&=\frac{\as(t)}{2\pi}
\gamma_{qq}^{\rm NS}\big(N,\as(t)\big) q_{\rm NS}(N,t).\cr}}
The operator matrix elements are related
to the corresponding parton distributions by Mellin transform,
i.e. the matrix element of the (leading twist) spin $N+1$ operator is
the $N$-th moment of the corresponding parton distribution
\eqn\mela{p(N,t)={\cal M}[p(x,t)]\equiv
\int_0^1\!dx\, x^N p(x,t),}
where $p(x,t)$ is any linear combination of parton densities.
Likewise, anomalous dimensions are found
by taking moments of the splitting functions
\eqn\melb{\gamma\big(N,\as(t)\big)=
{\cal M}[P(x,t)]\equiv\int_0^1\!dx\, x^N P(x,t).}
Notice that only operators for odd
values of $N>0$ in eq.~\mela,\melb\ actually exist,
because even (odd) spin operators are
charge-conjugation even (odd), while $F_2$ and the parton
distributions which contribute to it are charge-conjugation even;
the anomalous dimensions for all other (complex, in general) values of $N$
can only be  defined by analytic continuation. This continuation is
provided by the Altarelli-Parisi formalism, where the
primary quantities, namely the parton densities and splitting
functions, can be used to define the values of matrix elements and
anomalous dimensions for all $N$ through \mela\ and \melb, provided
only that they are known for all values of $0< x < 1$. Of course, these are
also precisely the quantities which can be extracted directly from
the structure functions measured experimentally, albeit in practice
only over a limited range of $x$.

The Altarelli-Parisi equations determine the parton distributions
at $(x',t')$ in terms of  their values
for all $x>x'$ and $t<t'$, hence they actually describe evolution
in the whole $(x,t)$ plane, even though the renormalization group
equations eq.~\rgeq\ for each value of $N$ only specify an
evolution law with respect to $t$. The evolution
with respect to $x$, while causal, is however nonlocal. The basic idea behind
double scaling is the realization that at small $x$ the Altarelli-Parisi
equations actually reduce to local evolution equations in both
variables, $x$ and $t$, which can then be treated symmetrically.

This can be shown by constructing a systematic approximation~\DAS\
to the evolution equations eq.~\aps\ at small $x$
and large $Q^2$. As $x$ gets smaller,
one would expect the behaviour of parton distributions to be dominated
by that of their Mellin transforms for
small $N$, and thus by the small $N$ behaviour of the anomalous dimensions.
Since the anomalous dimensions are singular at small $N$ one would
specifically expect their
rightmost singularity to provide the dominant behaviour.

A simple
way of showing~\DGPTWZ\ that this is indeed the case is to solve the
Altarelli-Parisi equations by Mellin transformation, i.e.
solve the renormalization group equations eq.~\rgeq\ for the
eigenvectors of the anomalous dimension matrix, which are
linear combinations $p(x,t)$ of the parton
densities. Using the leading order (LO) form of the anomalous
dimensions (which is $t$ independent) and of $\alpha_s=
\smallfrac{4\pi}{\beta_0 t}$ with $\beta_0=11-\smallfrac{2}{3}n_f$,
this procedure gives generically
\eqn\rgesol{p(N,t)=p(N,t_0)
\exp\left[{2\over\beta_0}\zeta\gamma^{(1)}(N) \right],}
where
\eqn\zdef{\zeta\equiv\ln\left({t\over t_0}\right)=\ln\left({
\ln(Q^2/\Lambda^2)\over\ln(Q^2_0/\Lambda^2)}\right).}
The $x$ space solution is then the inverse Mellin transform
\eqn\apesol{p(x,t)={\int_{-i\infty}^{i\infty}}\!dN\,\exp (\xi N)
p(N,t),}
where
\eqn\xidef{\xi\equiv\ln\left({x_0\over x}\right)}
and the integration runs over a contour located
to the right of all singularities
of $\gamma(N)$ and $p(N,t_0)$. Assume for the moment that any
singularities of the initial condition $p(N,t_0)$ are to the left of
those of $\gamma(N)$ (which are always poles on the real axis at
non-positive integer values of $N$).  Then at small $x$,
i.e. as $\xi$ grows the
integral may be evaluated by the saddle point method: the saddle point
condition is of the form
\eqn\saddle{
\xi_s+{2\over\beta_0}{d\gamma^{(1)}_{\rm sing}\over dN} \zeta=0,}
where $\gamma^{(1)}_{\rm sing}$ is the leading singularity of the one
loop anomalous dimension. Higher order terms in the expansion around
the singularity then give subleading corrections both to the location
of the saddle point and to the integral over it. If the initial condition
$q_{\rm NS}(N,t_0)$ has a singularity (typically a branch point)
to the right of that of the anomalous dimension this will also
contribute to the integral and indeed may dominate the contribution from
the saddle point as $\xi\to\infty$. However the saddle point always gives the
dominant contribution at large $Q^2$ i.e. large $\zeta$, and in any
case the dominant contribution to the evolution is always given at
large $\xi$ by the leading singularity $\gamma^{(1)}_{\rm sing}$.

We may thus obtain the leading small $x$ behaviour of the Altarelli-Parisi
equations by expanding the matrix of anomalous dimensions around its
rightmost singularity, determining the corresponding splitting
functions by inverse Mellin transformation, and then solving the
resulting simplified Altarelli-Parisi equations.
The singularity is located at $N=0$ in the singlet case, and at $N=-1$ in
the nonsinglet. This means that all other things being equal $q_{\rm NS}(x,t)$
will display the same qualitative behaviour as $x q_{\rm S}(x,t)$, and
$xg(x,t)$, i.e. at small $x$ the nonsinglet will be down by a power of
$x$ compared to the singlet.

Expanding the singlet anomalous dimension about its
leading singularity, at one loop we find
\eqn\smallnad{\gamma_{\rm S}^{(1)}(N)=
{1\over N}\pmatrix{2 C_A& 2C_F\cr0&0\cr}
+ \pmatrix{\smallfrac{11}{6}C_A-\smallfrac{2}{3}T_R n_f
&-\smallfrac{3}{2}C_F\cr
\smallfrac{4}{3}T_R n_f&0\cr}+O(N),}
while doing the same thing for the nonsinglet
\eqn\smallnadns{\gamma_{\rm NS}^{(1)}(N)={C_F\over {N+1}}+\half
C_F+O(N+1),}
where  $C_A = 3$, $C_F = \smallfrac{4}{3}$, $T_R = \half$ for QCD with
three colors. The corresponding splitting functions are
simply found by noting that
\eqn\melt{
{\cal M}\left[\smallfrac{1}{x}\right]=\smallfrac{1}{ N},\qquad
{\cal M}[\delta(1-x)]=1,\qquad {\cal M}\left[1\right]=\smallfrac{1}{N+1}.}
Positive powers of $N$ correspond to logarithmic derivatives of $\delta(1-x)$
which, when summed up, produce the nonlocal $x$ propagation kernel of eq.~\aps.
If instead the expansion is truncated, the evolution equations
are local. Specifically, using the splitting functions
obtained according to eq.~\melt\ from the anomalous dimensions eq.~\smallnad\
in the evolution equations eq.~\aps, and then differentiating with
respect to $\xi$, we get
\eqn\matweq{\eqalign{
{\partial\over\partial\xi\partial\zeta}&
\pmatrix{G(\xi,\zeta)\cr Q(\xi,\zeta)\cr}=\cr
&={2\over\beta_0}\left[\pmatrix{\smallfrac{11}{6}C_A-\smallfrac{2}{3}T_R n_f
&-\smallfrac{3}{2}C_F\cr
\smallfrac{4}{3}T_R n_f&0\cr}{\partial\over\partial\xi}+
\pmatrix{2 C_A& 2C_F\cr0&0\cr}\right]
\pmatrix{G(\xi,\zeta)\cr Q(\xi,\zeta)\cr},\cr
{\partial\over\partial\xi\partial\zeta}&\qns(\xi,\zeta)=
\frac{C_F}{\beta_0}\left[{\partial\over\partial\xi}+2\right]\qns(\xi,\zeta),\cr
}}
where we have for convenience defined
\eqn\eGQ{G(\xi,\zeta)\equiv xg(x,t),\qquad
         Q(\xi,\zeta)\equiv xq_{\rm S}(x,t),\qquad
  \qns(\xi,\zeta)\equiv q_{\rm NS}(x,t).}
The Altarelli-Parisi equations thus simply reduce to
two-dimensional partial differential equations with constant
coefficients, thereby proving the local nature of the evolution.

The singlet evolution equations eq.~\matweq\
can be solved by diagonalizing the matrix of anomalous dimensions
eq.~\smallnad: the eigenvectors will then satisfy (decoupled) equations
of the form eq.~\rgeq, with anomalous dimensions given
by the corresponding eigenvalues. To order $N^0$ the eigenvalues
are
\eqn\eval{\eqalign{&\lambda_+=2 C_A \smallfrac{1}{N} -
\left[\smallfrac{11}{6} C_A+\smallfrac{2}{3} T_R n_f- \smallfrac{4 C_F}{3 C_A}
T_R n_f\right]+O(N);\quad \lambda_-=- \smallfrac{4 C_F}{3 C_A}T_R n_f+O(N),}}
corresponding to  the eigenvectors  $v_\pm=(Q_\pm,G_\pm)$ given by
\eqn\evec{Q_+=\smallfrac{2 T_R}{3 C_A}n_f N G_+ +O(N^2)\qquad
Q_- =-\smallfrac{C_A}{C_F} G_- +O(N);}
the singlet quark and gluon distributions are then given by
$Q=Q_+ +Q_-$ and $G=G_+ +G_-$.
Notice that, according to standard perturbation theory,
when the eigenvalues
are determined up to next-to-leading order (in $N$),
only the leading nontrivial term ought to be kept in the
expression for the eigenvector.

Transforming to $x$ space
with the help of eq.~\melt\ we thus get
\eqn\weq{\eqalign{
\Big[\frac{\partial^2}{\partial\xi\partial\zeta}
+\delta_+\frac{\partial}{\partial\xi}-\gamma^2 \Big] G_+(\xi,\zeta)&=0,\cr
\Big[\frac{\partial}{\partial\zeta}+\delta_-\Big] G_-(\xi,\zeta)&=0,\cr
\Big[\frac{\partial^2}{\partial\xi\partial\zeta}
+\tilde\delta\frac{\partial}{\partial\xi}-\tilde\gamma^2 \Big]
\qns(\xi,\zeta)&=0,\cr}}
with
\eqn\dasparms{\gamma^2=\frac{12}{\beta_0},\quad
\tilde\gamma^2=\frac{8}{3\beta_0},\qquad\delta_+=
\frac{11+{2n_f\over 27}}{\beta_0},\quad\delta_-=
\frac{16n_f}{27\beta_0},\quad\tilde\delta=
\frac{4}{3\beta_0}.}
The eigenvector conditions eq.~\evec\ become
\eqn\xevec{Q_+(\xi,\zeta)=\frac{n_f}{9}\frac{\partial G_+(\xi,\zeta)}
{\partial\xi};
\qquad Q_-(\xi,\zeta)=-\frac{9}{4} G_-(\xi,\zeta).}
The eigenvector condition for $v_+$
can be conveniently rewritten by differentiating both sides
with respect to $\zeta$ and using the expression eq.~\weq\
for $\smallfrac{\partial^2 G_+}{\partial\xi\partial\zeta}$:
the term proportional to $\smallfrac{\partial G_+}{\partial\xi}$
can then be neglected since it only gives a subleading
correction to the eigenvector, and we get
\eqn\qeq{{\partial Q_+(\xi,\zeta)\over\partial\zeta}=
{n_f\over 9}\gamma^2 G_+(\xi,\zeta).}

Only the ``large'' eigenvalue $\lambda_+$ is singular at small $N$,
hence, at small $x$ and large $Q^2$,
the ``small'' eigenvalue may be neglected. The singlet quark and gluon
distributions are then no longer independent, but rather related
by the large eigenvector condition eq.~\xevec. If we had kept
only singular terms in the anomalous dimensions,  the small
eigenvalue would have vanished, the large eigenvector
would have coincided with the gluon, and the singlet quark distribution would
have vanished. This is also apparent by direct inspection of the
small $x$ evolution equation \weq, which would have coincided with
eq.~\matweq\ with all terms proportional to
$\partial\over\partial \xi$ on the \rhs\ neglected.
When the constant terms are retained, the large eigenvector contains a
mixture of gluons and singlet quarks,
obeying the evolution equation eq.~\weq\ for $G_+$ (and $Q_+$). This
has the same form as the gluon equation in eq.~\matweq, but with
the inhomogeneous term neglected, and a slightly different value of the
coefficient of
$\partial\over\partial \xi$. The singlet quark contribution $Q_+$ is
however subleading, being determined in terms of the gluon
contribution $G_+$ according to eq.~\qeq, which coincides
with the singlet quark equation in eq.~\weq.

It thus appears that the behaviour of the singlet and nonsinglet
components of $F_2$ at small $x$ and large $Q^2$ are
entirely determined by those of $G_+(\xi,\zeta)$ and $\qns(\xi,\zeta)$
respectively which, in turn, are determined
by the evolution equations eq.~\weq. These are recognized
as a two-dimensional wave equations, i.e.  two-dimensional Klein-Gordon
equations written in light-cone coordinates $(\xi,\zeta)=x\pm t$
with imaginary mass. This immediately implies several
general properties of their solution:
\item{(i)} The equations are essentially symmetrical in $\xi$ and $\zeta$, so
$G_+(\xi,\zeta)$ and $\qns(\xi,\zeta)$ evolve (`propagate') equally
in both $\xi$ and $\zeta$ (i.e. in $x$ and $Q^2$),
up to the (small) asymmetry induced by the `damping' term proportional
to $\delta$. Any further asymmetry in $\xi$ and $\zeta$ must thus come
{}~from the boundary conditions.
\item{(ii)} The propagation is `timelike', into the forward `light-cone'
at the origin $(\xi,\zeta)=(0,0)$, along the `characteristics'
$\xi={\rm constant}$ and $\zeta={\rm constant}$ (see fig.~2).
\item{(iii)} At a given point $(\xi,\zeta)$, $G_+(\xi,\zeta)$ and
$\qns(\xi,\zeta)$ depend
only on their respective boundary conditions contained within
the backward light cone formed by the two characteristics through
$(\xi,\zeta)$.
\item{(iv)} Because the equations are linear, contributions to
$G_+(\xi,\zeta)$ or $\qns(\xi,\zeta)$ from different parts
of the boundaries are simply added together.
\item{(v)} Since the `mass' terms are negative,
the propagation is `tachyonic'; this means that both $G_+(\xi,\zeta)$
and $\qns(\xi,\zeta)$ are unstable, growing exponentially rather
than oscillating.
\item{(vi)} Since $\delta$ and $\tilde\delta$ are both positive, the
damping terms ensure that, at fixed
$\xi$, $G_+(\xi,\zeta)$ and $\qns(\xi,\zeta)$ eventually fall
with increasing $\zeta$.

It is also straightforward to obtain the general solution to the
evolution equations: they are simple examples of the characteristic
Goursat problem, in which the solution
is entirely determined by the knowledge of boundary conditions along two
characteristics, and can be written explicitly in terms of a particular
solution to the equation. The latter is easily found by observing that,
setting  $z=2\gamma\sqrt{\xi\zeta}$, the first and third of eqns.~\weq\
coincide with the Bessel equation, the appropriate solution of which
is the Bessel function
\eqn\ebess{I_0(z)\equiv\sum_0^\infty\frac{\big(\quarter z^2\big)^n}{(n!)^2}
                 \toinf{z}\frac{1}{\sqrt{2\pi z}}e^z
                             \big(1+O(\smallfrac{1}{z})\big).}

The general solutions are thus found to be
\eqn\eGoursat{\eqalign{G_+(\xi,\zeta)&=
            I_0\big(2\gamma\sqrt{\xi\zeta}\big)e^{-\delta_+\zeta}G_+(0,0)
            +\int_0^\xi d\xi' I_0\big(2\gamma\sqrt{(\xi-\xi')\zeta}\big)
            e^{-\delta_+\zeta}\frac{\partial}{\partial\xi'}G_+(\xi',0)\cr
      &\qquad+\int_0^\zeta d\zeta'
I_0\big(2\gamma\sqrt{\xi(\zeta-\zeta')}\big)
                                   e^{\delta_+(\zeta'-\zeta)}
 \Big(\frac{\partial}{\partial\zeta'}
G_+(0,\zeta')+\delta_+G_+(0,\zeta')\Big),\cr
G_-(\xi,\zeta) &= e^{-\delta_-\zeta}G_-(\xi,0),\cr
\qns(\xi,\zeta) &=
            I_0\big(2\tilde\gamma\sqrt{\xi\zeta}\big)
                       e^{-\tilde\delta\zeta}\qns(0,0)
            +\int_0^\xi d\xi' I_0\big(2\tilde\gamma\sqrt{(\xi-\xi')\zeta}\big)
            e^{-\tilde\delta\zeta}\frac{\partial}{\partial\xi'}\qns(\xi',0)\cr
    &\qquad+\int_0^\zeta d\zeta'
I_0\big(2\tilde\gamma\sqrt{\xi(\zeta-\zeta')}\big)
                                   e^{\tilde\delta(\zeta'-\zeta)}
 \Big(\frac{\partial}{\partial\zeta'}\qns(0,\zeta')
                                +\tilde\delta \qns(0,\zeta')\Big).\cr}}
These solutions
display explicitly the symmetric nature of the evolution. It is
interesting to compare them with the standard solution to the Altarelli-Parisi
equations at larger $x$, where boundary conditions are imposed at a
scale $t_0$ for $x'$ such that $x\le x'\le 1$.  Since the structure function
vanishes kinematically at $x=1$ the boundary condition on the lower boundary
is trivial, and evolution takes place from the initial parton
distributions assigned at a given $t_0$ forwards in $t$.
In the present case, instead, the two boundaries are treated symmetrically,
and evolution takes place as much with respect to $x$ as it does with respect
to $t$.

The asymptotic behaviour of $G_+(\xi,\zeta)$ or $\qns(\xi,\zeta)$
at small $\xi$ and large $\zeta$,
i.e. far away from the boundary, will in general depend on the form of the
boundary conditions. Due to the linearity of the equation, we may consider
each boundary separately.
Contributions from each boundary are generated by fluctuations of the
functions $G_+$ or $\qns$ on that boundary. If these fluctuations
are sufficiently well localized close to the origin, then far from
the boundary we can use a multipole expansion, expanding the argument of the
Bessel
functions in powers of $\smallfrac{\xi^\prime}{\xi}$
(i.e., the distance from the boundary over the spread of the source)
for the left boundary
and  $\smallfrac{\zeta^\prime}{\zeta}$ for the lower boundary.
All the contributions from higher moments of the boundary fluctuations
are then seen to be suppressed
by powers of the light-cone distance from the origin
\eqn\sigmadef{\sigma\equiv\sqrt{\xi\zeta},}
while the leading contribution is simply given by the strength of the
source at the origin, and its asymptotic behaviour is determined by
that of the Bessel function \ebess:
\eqn\asymp{\eqalign{G_+(\rho,\sigma)&\neath\sim{\sigma\to\infty}
 {\cal N} \frac{1}{\sqrt{4\pi\gamma\sigma}}
       \exp\Big\{ {2\gamma\sigma -
      \delta_+\big(\smallfrac{\sigma}{\rho}\big)}\Big\}
      \big(1+O(\smallfrac{1}{\sigma})\big),\cr
  \qns(\rho,\sigma)&\neath\sim{\sigma\to\infty}
 \tilde{\cal N} \frac{1}{\sqrt{4\pi\tilde\gamma\sigma}}
       \exp\Big\{ {2\tilde\gamma\sigma -
       \tilde\delta\big(\smallfrac{\sigma}{\rho}\big)}\Big\}
      \big(1+O(\smallfrac{1}{\sigma})\big),\cr}}
where we have introduced the hyperbolic coordinate orthogonal to $\sigma$,
namely
\eqn\rhodef{\rho\equiv\sqrt{\xi/\zeta}}
(see fig.~2).\nref\GLR{L.V.~Gribov, E.M.~Levin and M.G.~Ryskin
\PRep\vyp{100}{1983}{1}.}
\nfig\fxize{The ($\xi$,$\zeta$) plane,
showing the backward light cone
          at the point ($\xi'$,$\zeta'$), curves of constant
          $\sigma$ (the hyperbolae) and lines of constant $\rho$.
The origin is chosen in such a way that the small $x$ approximation
of the evolution equations is valid for positive $\xi$, and perturbation
theory breaks down for negative $\zeta$ where $\alpha_s$ grows too large.
The hatched area  $x\lsim x_r\exp(-\alpha_s(t_0)^2/\alpha_s(t)^2)$
indicates the region where parton recombination effects are
expected~\GLR\ to lead to breakdown of simple perturbative evolution due to
higher twist corrections.}

The origin of  double scaling is now clear: because
of the isotropy of the evolution equation, its solution asymptotically
only depends on the scaling variable $\sigma$, i.e., its level
curves are hyperbolae
in the $(\xi,\zeta)$ plane (see fig.~2), and do not depend on the
direction in which the propagation occurs ($\rho$ scaling).
Furthermore, because the solutions \asymp\ are
asymptotically independent of the boundary
conditions, the dependence on $\sigma$ is given by a universal
rise, stronger than any power of $\xi$ but weaker than any inverse
power of $x$. The universal form of this rise
reflects the underlying dynamical mechanism which generates
it, namely, in the singlet case the collinear singularity in the
triple gluon vertex
which is responsible for the singularity in the anomalous
dimension $\gamma_{gg}$ (eq.~\smallnad) whose strength determines the
coefficient $\gamma^2$ (eq.~\dasparms), and in the nonsinglet the
corresponding singularity in $\gamma_{qq}^{\rm NS}$ due to collinear
gluon bremsstrahlung, whose strength determines the coefficient
$\tilde\gamma^2$. In an abelian theory the former singularity
clearly vanishes, so the singlet distributions would no longer
grow so strongly, behaving instead more like the nonsinglet.

Using eq.~\qeq\ to determine $Q$, and the expression eq.~\ftsx\ of $F_2$,
and neglecting the nonsinglet contribution, since this is suppressed
by a factor of approximately $e^{-\xi}$, it is easy to show that the
asymptotic behaviour of $F_2^p$ (or $F_2^n$) will also have a double
scaling form, namely:
\eqn\eFasymp{F_2^p\neath\sim{\sigma\to\infty}
    \frac{5n_f}{162}{\gamma\over\rho}
    G_+(\sigma,\rho)\big(1+O(\smallfrac{1}{\sigma})
                         +O(\smallfrac{1}{\rho})\big),}
with $G_+$ given by eq.\asymp.
This behaviour holds up to corrections of order $1\over\sigma$,
{}~from the subasymptotic form of the Bessel function and the boundary
corrections, up to terms of relative order $\sigma\over \rho^2$ in
the exponent,
{}~from higher order contributions to the small $N$ expansion eq.~\smallnad\
of the anomalous dimensions, and up to corrections of order $1\over \rho$,
{}~from higher order contributions to the eigenvector equation. It thus
holds in the  limit $\sigma\to\infty$ along any
curve such that also $\rho\to\infty$, such as
for example the curve $\xi\propto \zeta^{1+\epsilon}$ with $\epsilon>0$:
that is, far from the boundaries, and provided the increase
of $\ln{1\over x}$ is more rapid than that of $\ln t$.

All of this, however, hinges on the assumption that the solution
eq.~\eGoursat\ to the wave equation may be treated in the multipole
expansion, i.e., that the fluctuations on the boundaries fall off away
>from the origin.
If this does not happen, then the multipole expansion is not valid, the
asymptotic behaviour eq.~\asymp\ does not hold, and the chain of arguments
leading to the scaling form of $F_2$ eq.~\eFasymp breaks down.
Specifically, assume for example that the boundary conditions at $\zeta=0$
are exponentially rising functions of $\xi$:
$G_+(\xi,0)\sim \exp\lambda\xi=x^{-\lambda}$, $\qns(\xi,0)\sim
\exp\tilde\lambda\xi=x^{-\tilde\lambda}$. The boundary integral may then
be evaluated by the saddle point method, is dominated by a nontrivial
saddle-point,
and gives the asymptotic behaviour
\eqn\ehpa{\eqalign{G_+(\sigma,\rho)&\toinf{\sigma} {\cal N}^\prime
\exp{\big\{\lambda\sigma\rho +
(\smallfrac{\gamma^2}{\lambda}-\delta_+)(\smallfrac{\sigma}{\rho})\big\}}
      \big(1+O(\smallfrac{1}{\sigma})+O(\smallfrac{1}{\rho})\big),\cr
\qns(\sigma,\rho)&\toinf{\sigma} \tilde{\cal N}^\prime
\exp{\big\{\tilde\lambda\sigma\rho +
(\smallfrac{\tilde\gamma^2}{\tilde\lambda}-\tilde\delta)
                          (\smallfrac{\sigma}{\rho})\big\}}
      \big(1+O(\smallfrac{1}{\sigma})+O(\smallfrac{1}{\rho})\big).\cr}}
The strong growth on the boundary is thus preserved by the evolution.
Since the evolution equations are linear, this behaviour should be
added to the dynamically generated contributions \asymp. Since it
is powerlike, the rise of the boundary condition will eventually dominate the
dynamically generated rise eq.~\asymp\ when $\xi$ is large enough.
However the universal behaviour eq.~\asymp\ is still dominant when $\zeta$
is large enough that the nontrivial saddle point leading to
eq.~\ehpa\ is no longer dominant, i.e. whenever
$\rho\lsim{\gamma\over\lambda}$,
$\rho\lsim{\tilde\gamma\over\tilde\lambda}$ respectively. Similar
contributions to \ehpa\ would arise from exponentially rising
functions of $\zeta$ on the lower boundaries.

In practice, we may choose $x_0$ close to the turning point of the
evolution, so that the lower boundary condition is reasonably flat.
Furthermore Regge theory suggests that the left hand boundary
condition is given by $F_2\sim x^{1-\alpha(0)}$, where $\alpha(0)$ is
the intercept of the appropriate Regge trajectory. In the singlet
channel, this is the pomeron trajectory, with~\landrev\ $\alpha_P(0)\simeq
1.08$, so the left hand boundary condition for $G_+$ should also be
soft, and the asymptotic behaviour of $F_2^p$ and $F_2^n$ given by
\eFasymp, at least for $\rho\lsim\gamma/(\alpha_P(0)-1)\simeq 15$.
However in the nonsinglet channel, the appropriate trajectory is that
of the $\rho$, and $\alpha_\rho(0)\simeq\half$. So the left hand
boundary condition for the nonsinglet, $\qns\sim
x^{-\alpha_\rho}$, is hard, and asymptotically \asymp\ will be
dominated by \ehpa, whence, for $\rho\gsim,
\tilde\gamma/\tilde\lambda\simeq 1$
\eqn\eFasympns{F_2^p-F_2^n \neath\sim{\sigma\to\infty}
    \tilde{\cal N}''\exp{\big\{(\tilde\lambda-1)\sigma\rho +
(\smallfrac{\tilde\gamma^2}{\tilde\lambda}-\tilde\delta)
                          (\smallfrac{\sigma}{\rho})\big\}}
      \big(1+O(\smallfrac{1}{\sigma})+O(\smallfrac{1}{\rho})\big),
}
with $\tilde\lambda\simeq\half$. Recent data from NMC~\ref\NMC{
NMC Collaboration, M.~Arneodo et al., \PR\vyp{D50}{1994}{1}.} seem
to be in good qualitative agreement with this prediction, but deuteron
data in the kinematic region explored at HERA would be necessary to
confirm it more precisely.

It was suggested some time ago~\ref\KMRS{J.~Kwieci\'nski et al.,
\PR \vyp{D42}{1990}{3645}.} that the singlet boundary condition at $Q_0^2$
might not be given by the intercept of the pomeron trajectory, but
could rather rise very steeply as $x^{-\lambda_L}$, with
$\lambda_L=4\ln 2 \smallfrac{C_A}{\pi}\as\simeq\half$. This steep
initial rise (sometimes called the `hard pomeron') was supposed to
incorporate, in an admittedly rather heuristic manner, the higher
order perturbative effects at small $x$
described by the BFKL equation~\ref\BFKL{
       L.N.~Lipatov, \SJNP\vyp{23}{1976}{338}\semi
          V.S.~Fadin, E.A.~Kuraev and L.N.~Lipatov,
       \PL\vyp{60B}{1975}{50};
       {\it Zh.E.T.F.~}\vyp{71}{1976}{840}, \vyp{72}{1977}{377};
       {\it Sov. Phys. JETP~}\vyp{44}{1976}{443};\vyp{45}{1977}{199}\semi
          Y.Y.~Balitski and L.N.Lipatov, \SJNP\vyp{28}{1978}{822}\semi
      For a pedagogical introduction see V.~Del~Duca, preprint DESY 95-023,
       {\tt hep-ph/9503226}.}.
With such a boundary condition, the double scaling
rise \eFasymp\ in $F_2^p$
would be masked by the stronger rise of $G_+$ as given by \ehpa, so
\eqn\eFasympLip{F_2^p\neath\sim{\sigma\to\infty}
    {\cal N}''\exp{\big\{\lambda_L\sigma\rho +
(\smallfrac{\gamma^2}{\lambda_L}-\delta)
                          (\smallfrac{\sigma}{\rho})\big\}}
      \big(1+O(\smallfrac{1}{\sigma})+O(\smallfrac{1}{\rho})\big),}
whenever $\rho\gsim \gamma/\lambda_L \simeq 2$. This prediction is
nonuniversal, in the sense that the precise slope of the rise cannot
be predicted since $\lambda_L$ depends on $\as$, and it is not known
at which scale $\as$ should be determined. Furthermore it is
qualitatively different in form from
the double scaling rise \eFasymp: in particular the rise at large
$\xi$, fixed $\zeta$ is now no longer accompanied by a corresponding rise
at large $\zeta$ and fixed (though large) $\xi$.
We will come back on the issue of applicability of
leading order computations and the relative importance of higher order
corrections in the next section. First however we will see whether the HERA
data support the universal double scaling prediction \eFasymp, or prefer the
more phenomenological suggestion \eFasympLip.

The scaling plots in fig.~1 display the measured values of $F_2$,
with the subasymptotic corrections in eq.~\eFasymp\ rescaled out, i.e.
$R_F'F_2$ with
\eqn\RFp{R_F'(\sigma,\rho)=\exp\big(\delta(\sigma/\rho)
+\half\ln\gamma\sigma+\ln(\rho/\gamma)\big).}
The scaling variables are computed with $x_0=0.1$, $Q_0=1$~GeV,
$\Lambda=263$~MeV, and $\delta=\delta_+$ eq.~\dasparms\ with $n_f=4$.
The data cover a wide span in $\rho$: for instance the data in fig.~1c
have $1\lsim \rho\lsim 5$. Nevertheless they all fall on the same line,
and display a slope which agrees very well with the predicted asymptotic
value $2\gamma=2.4$. If the boundary condition were hard,
 eq.~\ehpa\ shows that the leading behaviour would
also be a linear rise of $\ln F_2$ in $\sigma$, but now with a slope which
is not universal (as it depends on $\lambda$), and strongly $\rho$ dependent.
Hence, the data should not fall on a single line, and
the agreement of the observed slope with the calculated value of $2\gamma$
could only be a coincidence. The fact that the data display double scaling
thus allows us to exclude the possibility of
power-like boundary conditions to leading-order (or, as we will se in
a moment, next-to-leading order) evolution at a very high confidence
level~\test.

The scaling plots in fig.~1 also show that the slope of
the rise of $\ln F_2$ is significantly
smaller than the asymptotic one when $\sigma$ is not too large.
This suggests that
scaling violations may already be important here. Even more dramatic
scaling violations are seen if one considers data at low $Q^2$.
Both effects are illustrated
in fig.~3~(i),  which displays $F_2$ after complete rescaling
of the leading asymptotic behaviour, i.e. by a factor
\eqn\RF{R_F(\sigma,\rho)=\exp\big(-2\gamma\sigma + \delta(\sigma/\rho)
+\half\ln\gamma\sigma+\ln(\rho/\gamma)\big).}
It is apparent that the data display a systematic drop in $\sigma$
(the asymptotic double scaling line is approached from above),
and the recent (albeit preliminary) data with low
$Q^2$ ($2\lsim\rho\lsim 5$ and $1.3\lsim\sigma\lsim1.8$) do not
seem to scale at all.\nref\shifvert{ZEUS Collaboration, DESY 95-193,
{\tt hep-ex/9510009}.}\nfig\scalviol{Double scaling
plots of $R_F F_2^p$ against i) $\sigma$ and ii) $\rho$, (a) with
rescaling of the LO asymptotic behaviour eq.~\RF;
(b) with NLO rescaling eq.~\RFtwo. The stars are preliminary low $Q^2$
($1.5\GeV^2\le Q^2\le 15\GeV^2$, $3.5~ 10^{-5}<x<4.0~10^{-3}$) points from
ZEUS~\shifvert.}

This leads us to consider scaling violations, the simplest of which appear
when two-loop corrections are included in the Altarelli-Parisi
equations. The leading singularities of the two loop anomalous
dimensions~\ref\twoloop{E.G.~Floratos \etal, \NP\vyp{B192}{1981}{417}.}
in \MS\ scheme are
\eqn\twoloopsing{\eqalign{
\gamma_{\rm S}^{(2)}(N)&={1\over N}\pmatrix{
(\smallfrac{4}{3}C_F-\smallfrac{46}{9}C_A) T_R n_f
& C_F C_A-\smallfrac{40}{9}C_F T_R n_f\cr
\smallfrac{40}{9}C_A T_R n_f&\smallfrac{40}{9}C_F T_R n_f\cr}
+ O(1),\cr
\gamma_{\rm NS}^{(2)}(N)&={C_F^2\over {(N+1)^3}}+O((N+1)^{-2}).\cr}}
Note that all the entries in the two loop singlet anomalous dimension
are singular, and the nonsinglet is more singular than at one loop.
Taking the inverse Mellin transform of \twoloopsing, using for the
nonsinglet the result ${\cal M}\left[\half{1\over x} \ln^{2}{1\over x}
\right]=1/N^3$, and including them in the Altarelli-Parisi equations
\aps\ gives wave equations similar to \matweq, but with additional
terms of $O(\as)$ on the \rhs.
Linearizing these two loop corrections, the wave equations \weq\
become (suppressing subasymptotic contributions)
\eqn\weqtwo{\eqalign{
\Big[\frac{\partial^2}{\partial\xi\partial\zeta}
+\delta_+\frac{\partial}{\partial\xi}-\gamma^2 \Big]
G_+(\xi,\zeta)&=\epsilon_+ \as(t_0)e^{-\zeta}G_+(\xi,\zeta),\cr
\Big[\frac{\partial}{\partial\zeta}+\delta_-\Big] G_-(\xi,\zeta)
&=\epsilon_- \as(t_0)e^{-\zeta}G_+(\xi,\zeta),\cr
\Big[\frac{\partial^2}{\partial\xi\partial\zeta}
+\tilde\delta\frac{\partial}{\partial\xi}-\tilde\gamma^2 \Big]
\qns(\xi,\zeta)&=\tilde\epsilon\as(t_0)e^{-\zeta}\int_0^\xi
\!d\xi^\prime\,(\xi-\xi^\prime)\qns(\xi^\prime,\zeta),\cr}}
with now, in place of \zdef,
\eqn\zdeftwo{\zeta\equiv\ln\left({\as(t_0)\over \as(t)}\right),}
with $\as$ evaluated to two loops, and
\eqn\dastwoparms{\epsilon_+=
\Big(\frac{103}{27}n_f+3\frac{\beta_1}{\beta_0}\Big)\Big/\pi\beta_0,\quad
\epsilon_-=\frac{26n_f}{3\pi\beta_0},
\quad\tilde\epsilon=\frac{16}{3\pi\beta_0},}
$\beta_1=102-\smallfrac{38}{3}n_f$ being the two loop coefficient of the
$\beta$-function. It is not difficult to show that the three new
parameters \dastwoparms\ are in fact all independent of the
choice of factorization scheme. The eigenvector conditions
\evec,\xevec\ and \qeq\ are unchanged.

The general solutions of the three equations \weqtwo\ are the same as
those of the leading order equations, \eGoursat, but each with an
additional contribution on the \rhs:
\eqn\eGtwo{\eqalign{
&\epsilon_+\frac{4\pi}{\beta_0}\int_0^\xi\int_0^\zeta\!d\xi'd\zeta'
I_0\big(2\gamma\sqrt{(\xi-\xi')(\zeta-\zeta')}\big)
e^{\delta_+(\zeta'-\zeta)-\zeta'}
G_+(\xi',\zeta'),\cr
&\epsilon_-\frac{4\pi}{\beta_0}\int_0^\zeta\!d\zeta'
e^{\delta_-(\zeta'-\zeta)-\zeta'}
G_+(\xi',\zeta'),\cr
&\tilde\epsilon\frac{4\pi}{\beta_0}
\int_0^\xi\int_0^{\xi'} \int_0^\zeta\!d\xi'd\xi''d\zeta'
I_0\big(2\gamma\sqrt{(\xi-\xi')(\zeta-\zeta')}\big)
e^{\delta_+(\zeta'-\zeta)-\zeta'}(\xi'-\xi'')\qns(\xi'',\zeta'),\cr}}
respectively. Two loop corrections to the leading asymptotic behaviour
are then found by substituting the leading behaviour into \eGtwo, and
evaluating the asymptotic form of the integrals. Clearly the
form of the correction will then depend on the form of the leading
behaviour. For soft boundary conditions the double scaling behaviours
\asymp\ are corrected by extra factors
\eqn\asymptwo{\eqalign{G_+(\rho,\sigma)&\neath\sim{\sigma\to\infty}
{\cal N} \frac{1}{\sqrt{4\pi\gamma\sigma}}
\exp\Big\{ {2\gamma\sigma -
      \delta_+\big(\smallfrac{\sigma}{\rho}\big)}\Big\}
\Big[1-\epsilon_+\big(\as(t_0)-\as(t)\big)\smallfrac{\rho}{\gamma}\Big]
      \big(1+O(\smallfrac{1}{\sigma})\big)
,\cr
\qns(\rho,\sigma)&\neath\sim{\sigma\to\infty}
 \tilde{\cal N} \frac{1}{\sqrt{4\pi\tilde\gamma\sigma}}
       \exp\Big\{ {2\tilde\gamma\sigma -
       \tilde\delta\big(\smallfrac{\sigma}{\rho}\big)}\Big\}
\Big[1-\tilde\epsilon\big(\as(t_0)-\as(t)\big)
\smallfrac{\rho^3}{\tilde\gamma^3}\Big]
      \big(1+O(\smallfrac{1}{\sigma})\big)
,\cr}}
while the double scaling behaviour of $F_2$, \eFasymp, is corrected by
a similar factor:
\eqn\eFasymptwo{\eqalign{F_2^p\neath\sim{\sigma\to\infty}
    &\frac{5n_f}{162}{\gamma\over\rho}
{\cal N} \frac{1}{\sqrt{4\pi\gamma\sigma}}
\exp\Big\{ {2\gamma\sigma -
      \delta_+\big(\smallfrac{\sigma}{\rho}\big)}\Big\}\cr
&\qquad\Big[1-\big(\epsilon_+\big(\as(t_0)-\as(t)\big)
-\smallfrac{9\epsilon_-}{n_f\gamma^2}\as(t)\big)
\smallfrac{\rho}{\gamma}\Big]
\big(1+O(\smallfrac{1}{\sigma})+O(\smallfrac{1}{\rho})\big),\cr}}
the extra term in the square brackets coming from the
subleading contribution to $G_-$.
For hard boundary conditions, the asymptotic behaviours \ehpa,
\eFasympns, are corrected by the same factors \asymptwo, \eFasymptwo\
respectively, but with $\rho/\gamma$, $\rho/\tilde\gamma$ replaced
by $\lambda$, $\lambda$ respectively: the corrections are thus then
$x$ independent.

\nref\blois{R.D.~Ball and S.~Forte, CERN-TH.7421/94, {\tt hep-ph/9409373},
        to be published in {\it ``The Heart of the Matter''},
        VIth Rencontre de Blois, June 1994 (Editions Fronti\`eres).}
\nref\mont{R.D.~Ball and S.~Forte,
   in the proceedings of {\it ``QCD94''}, Montpellier, July 1994
   (\NPBPS\vyp{39B,C}{1995}{25}).}
\nref\alphas{R.D.~Ball and S.~Forte, \PL\vyp{B358}{1995}{365}.}
Two loop corrections will be most important for soft
boundary conditions, and thus in the singlet channel, as the leading
correction to the double scaling seen at HERA. To see their effect,
the data are replotted [in fig.~3~(ii)] with a new rescaling function,
\eqn\RFtwo{R_F^{(2)}(\sigma,\rho)=R_F(\sigma,\rho)
\Big[1-\as(t_0)\Big(\epsilon_+ - (\epsilon_+ +
\smallfrac{9\epsilon_-}{n_f\gamma^2})e^{-\sigma/\rho}\Big)
\smallfrac{\rho}{\gamma}\Big]\inv,}
where $R_F$ is the leading order rescaling \RF. It can be seen from
the plots that the effect of the two loop correction is moderate
(except at very low $Q^2$) but
significant in the range of the present data: it increases the
starting scale from $Q_0\simeq 1~\GeV$ at leading order to
around $Q_0\simeq 1.5~\GeV$, reduces the slope of the $\sigma$-plot by
about $10\%$, and decreases the rise in the $\rho$-plot at large
$\rho$ and low $Q^2$. Thus most of the as yet observed scaling
violations can be accounted for by the two loop correction, and
conversely the effect of this correction can be clearly seen in the data.
These results have been confirmed by numerical calculation using the
full one and two loop anomalous dimensions~\refs{\blois-\alphas}.

The residual rise at large $\rho$ in the low $Q^2$ data \shifvert, if
it turns out to be statistically significant, could be due
to many different (and possibly competing) nonperturbative effects:
a small rise in
the (nonperturbative) boundary condition (even $x^{-0.08}$ has
observable effects at $Q_0=1.5\GeV$), nonperturbative effects due to
the opening of the charm threshold, conventional higher twist effects,
or even more novel higher twist effects such as parton recombination.
However it could also be due to higher loop singularities.
It is particularly important to consider these, since it is necessary,
in the light of the above discussion about the `hard pomeron' boundary
condition above, to understand why they do not in fact
spoil double scaling. Such an undertaking is possible since the
precise form of the leading (and some of the sub-leading)
singularities are known: we will now explain in some detail how their
effects may be properly included.

\bigskip
\newsec{Perturbation Theory at Small $x$}
\medskip
In the previous section we have seen how double scaling appears
as a generic feature of the solution to the LO or NLO
Altarelli-Parisi equations, in the limit as $\rho$ and $\sigma$
grow large, i.e. as both $\ln \smallfrac{1}{x}$ and
$\ln t$ grow, provided the former grows faster than the
latter. This prediction thus defines a ``double scaling limit'' of QCD,
intermediate between the Regge limit (small $x$ at fixed $t$),
and the Bjorken limit (large $t$ at fixed $x$). In the Regge limit
perturbation theory fails and we are unable to calculate either the
$x$ or the $t$ dependence of parton distributions. In the Bjorken
limit perturbation theory holds and predicts the scale dependence
of parton distributions, viewed as functions of $x$, but the $x$
dependence itself depends on an uncalculable initial condition. In
the double scaling limit, the $x$ and $t$ dependence is universal and
only depends on a single overall normalization.

We can thus divide
the $(\xi,\zeta)$ plane in various regions (see fig.~2): for low $\zeta$ (say,
$\zeta<0$ with a suitable choice the origin of coordinates) perturbation theory
breaks down, whereas the Altarelli-Parisi equations
hold for positive values of $\zeta$, and become more and more accurate
as $\zeta$ increases. On the other hand if $\xi$ is also sufficiently
large, the anomalous dimensions may be expanded around their leading
singularities eq.~\smallnad\ so the Altarelli-Parisi
equations take the small $x$ form eq.~\matweq.  However if
$\xi$  keeps increasing, perturbation theory eventually breaks down
because higher twist corrections, necessary to ensure unitarity,
must become get more important. Physically,
one can understand this by noting that $x$ can be interpreted as the
momentum fraction carried by individual partons. When $\xi$ is very
large momentum is shared between an increasingly large number of partons;
this corresponds to an increase of the parton density which cannot continue
indefinitely and should eventually stop when partons start
to recombine with each other~\GLR. This is expected to
happen in the region  $x\lsim x_r\exp(-\alpha_s(t_0)^2/\alpha_s(t)^2)$,
where $x_r$ cannot be reliably computed but could be of order $10^{-5}$
at scales of a few GeV$^2$~\GLR. If instead $\zeta$ keeps increasing
we eventually get to a region where effects from the lower boundary of
perturbative small $x$ evolution are important: these reflect
the shape of parton distributions propagated down from large values of $x$,
and thus we get back to Bjorken scaling, with uncalculable dependence
on $x$.

Even in the remaining double scaling region, however, double scaling
actually holds only with sufficiently soft boundary conditions; besides,
we have derived it only in a LO or NLO calculation. The two issues
are actually closely related, because of general arguments suggesting that
when higher order corrections are included a hard power-like rise of
the singlet parton distributions may result. The underlying
logic is the following: the leading logarithmic gluon-induced
contributions to the deep-inelastic scattering cross section
to all orders in $\alpha_s \log{1\over x}$ may be summed up by
the solution of an equation satisfied by the gluon distribution
(BFKL equation)~\BFKL. This
equation is not consistent with the renormalization group, in that
it does not sum logs of $Q^2$ and thus it does not include evolution in $Q^2$:
in fact, it is derived at fixed coupling $\alpha_s$. It therefore also
does not separate leading twist from higher twist contributions.
However, it is possible ~\ref\jar{T.~Jaroszewicz, \PL \vyp{116B}{1982}{291}.}
to extract the leading twist contribution to
the solution of this equation in Mellin space: as $Q^2\to\infty$
the solution has the form
\eqn\bfklsol{G(N,Q^2)= \left({Q^2\over Q_0^2}
\right)^{\gamma(N,\alpha_s)}G(N,Q^2_0)\left[1+O\left({Q_0^2\over Q^2}
\right)\right],}
the extra term in the square brackets being higher twist. The leading
twist piece is now the same as a solution to a (fixed coupling)
renormalization group equation eq.~\rgeq, with $\gamma(N,\as)$
identified as the anomalous dimension.
This anomalous dimension is determined as the
inverse of the function $\chi(x)=2\psi(1)-\psi(x)-\psi(1-x)$ (where
$\psi(x)$ is the Euler function):
\eqn\recrel{
\chi\left[\gamma\left(N, \alpha_s\right)\right]={N\over \bar\as},}
where $\bar\as\equiv{C_A\alpha_s\over\pi}$.

The solution to \recrel, the Lipatov anomalous dimension $\gamma_L$,
thus turns out to be a function of $\bar\alpha\over N$:
\eqn\bfklexp{\gamma_{L}\left(\bar\as/N\right)=\sum_{k=1}^N
\gamma_{L}^{(k)}\left({\bar\alpha\over N}\right)^k,}
with coefficients $\gamma_{L}^{(k)}$ determined uniquely by
eq.~\recrel. All the coefficients turn out to be positive, save
$\gamma_{L}^{(2)}$, $\gamma_{L}^{(3)}$ and $\gamma_{L}^{(5)}$ which
vanish. Now, inverse powers of $N$ are the Mellin transforms
of logs of $1\over x$:
\eqn\melsing{{\cal M}\left[{1\over x}{1\over (k-1)!} \ln^{k-1}{1\over x}
\right]={1\over N^k}.}
One may then argue~\jar\ that, since
the leading
logs of $1\over x$ are summed by the BFKL equation which
leads to eq.~\bfklsol, the coefficients $\gamma_{L}^{(k)}$
in eq.~\bfklexp\ must give
the coefficient of the most singular term in  expansion in powers of $N$
of the $k$-th order
contribution to the ordinary anomalous dimension (leading singularity):
these are by definition the leading logs in $1\over x$ since in $x$
space they correspond to the contributions with the largest number of logs
at each perturbative order.
This result may actually be proven rigorously
by means of suitable factorization theorems~\ref\fact{S.~Catani,
F.~Fiorani and G.~Marchesini,
\PL\vyp{B234}{1990}{339}; \NP\vyp{B336}{1990}{18}\semi
  S.~Catani, F.~Fiorani, G.~Marchesini and G.~Oriani,
\NP\vyp{B361}{1991}{645}\semi
S.~Catani, M.~Ciafaloni and F.~Hautmann,
            \PL\vyp{B242}{1990}{97}; \NP\vyp{B366}{1991}{135}.}:
the  expansion of
the Lipatov anomalous dimension eq.~\bfklexp\ gives the coefficients
of the leading singularities to all orders in $\alpha_s$
in the gluon anomalous dimension $\gamma_{gg}$
(due to the fact that the BFKL equation describes gluon propagation
and emission). Notice that these
coefficients are factorization scheme independent.

This result has several important consequences for our discussion. First,
it shows explicitly that the double logarithms, of the form
$\smallfrac{1}{x}\as(\as\ln^2\smallfrac{1}{x})^{n-1}$, which one might
expect naively to arise in a perturbative expansion of splitting
functions ~\ref\Gross
{D.~Gross in the proceedings of the XVII Intern. Conf. on
High Energy Physics, London, 1974
                       (published by SRC, Rutherford Lab.) and
lectures given at Les Houches, Session XXVIII, 1975, published in
``Methods in Field Theory'', ed. R.~Balian and J.~Zinn-Justin
(North-Holland, 1976).}, are reduced in the gluon sector to
single logarithms of the form
$\smallfrac{1}{x}\as(\as\ln\smallfrac{1}{x})^{n-1}$, because many of
the singularities cancel systematically. This cancellation actually
occurs in the whole singlet sector~\ref\singfact{
S.~Catani, M.~Ciafaloni and F.~Hautmann,\PL\vyp{B307}{1993}{147}.}
(though not in the nonsinglet~\ref\lipkir{R.~Kirschner and L.~Lipatov,
\NP\vyp{B213}{1983}{122}.}).
This is in accordance with the calculation of the two loop
singularities \twoloopsing\ (which in fact exhibit a further accidental
cancellation in the singlet sector, which also occurs at
three and five loops).

Despite this remarkable cancellation, at higher orders
in $\alpha_s$ the singularity in the
anomalous dimension is still growing strongly, albeit not quite so
fast as one might have naively expected. Thus at small
enough $x$ the enhancement
due to the extra powers of $\ln{1\over x}$ in the corresponding splitting
function may offset the suppression due to the extra powers of $\alpha_s$,
so that the inclusion of higher order corrections may be required in
order to obtain accurate results. Furthermore, the function
$\chi(x)$ has a symmetric minimum at $x=1/2$,
implying that the anomalous dimension $\gamma_{L}(\bar\as/N)$
has a square-root branch point there (there are also other branch
points, and in fact the structure of $\gamma_{L}(\bar\as/N)$ in the
complex plane is quite complicated~\ref\ehw{R.K.~Ellis,
F.~Hautmann and B.R.~Webber, \PL\vyp{B348}{1995}{582}.}). The value
of its argument such that $\gamma_{L}(\bar\as/N)=1/2$ is $4\ln 2$, so
the branch point is at
\eqn\lipex{
\lambda_L(\alpha)\equiv4\ln 2 \bigg(\frac{C_A\alpha}{\pi}\bigg).}
This implies that Mellin-space parton distributions also have a
singularity at $N=\lambda_L$, and correspondingly, in $x$ space they should
increase as $x\to 0$ as $x^{\lambda_L}$. This corresponds to a rather
strong rise
for realistic values of $\alpha_s$, and would certainly spoil
the observed double scaling behaviour if it occurred already at
presently attainable values of $x$. However, it is not
clear that this simple argument is consistent with the renormalization
group: for example, the result
of using a power like behaviour of this form as a boundary condition
for LO or NLO perturbative evolution depends on the scale $Q_0$
at which the boundary is set. Moreover, it is not clear how this
behaviour should be matched to conventional perturbative
evolution at larger $x$, nor indeed in which region of the $x$-$t$
plane it sould become important.

We need thus to keep into account higher order singularities in $N$
in a way consistent with the renormalization
group. This can be
done~\ref\summ{R.D.~Ball and S.~Forte, \PL\vyp{B351}{1995}{313}.}
by reorganizing the perturbative expansion of the anomalous
dimensions, or, equivalently, the splitting functions
used in the Altarelli--Parisi equations. To understand how this
works, it is convenient to classify the contributions to anomalous
dimensions by expanding the anomalous dimension used
in renormalization group equations eq.~\rgeq\ in powers of $\alpha_s$,
and then each order in powers of $N$
(see fig.~4):
\eqn\laxexpad
{{\alpha_s\over 2\pi}\gamma(N,\alpha)= \sum_{m=1}^\infty \alpha^m
\sum_{n=-\infty}^m A^m_n N^{-n}
=\sum_{m=1}^\infty \alpha^m
\bigg(\sum_{n=1}^m A^m_n N^{-n} + \bar\gamma_N^{(m)}\bigg),}
where the  numerical coefficients $A^m_n$ are given by eq.~\bfklexp,
and in the last step we have
separated out the regular part of the anomalous dimension
$\bar\gamma_N^{(m)}$.
Using eqs.~\melt,\melsing\ this is seen to corresponds to
expanding the associated splitting functions as
\eqn\laxexpsf
{{\alpha_s\over 2\pi}P(x,t)=\sum_{m=1}^\infty \big(\alpha_s(t)\big)^m
\bigg(\frac{1}{x}\sum_{n=1}^m A^m_n\frac{\ln^{n-1}\smallfrac{1}{x}}{(n-1)!}
+ \bar P^{(m)}(x)\bigg),}
where $\bar P^{(m)}(x)$ are regular as
$x\to 0$.

Solving renormalization group equations sums all leading
logs of the scale which appears in the equation: for instance, upon solving
eq.~\rgeq, the anomalous dimension gets
exponentiated according to eq.~\rgesol. At LO only the term with $m=1$
is included in the anomalous dimension eq.~\laxexpad, which is thus linear
 in $\alpha_s$, so  this amounts to
summing up all contributions
to the deep-inelastic cross sections where each extra power
of $\alpha_s$ is accompanied by a power of $\ln Q^2$. In fact, because the
LO anomalous dimension has a $1\over x$ singularity (which leads to a
factor of $\ln{1\over x}$ in the cross section upon integration)
some logs of ${1\over x}$  are also summed, but
$\ln {1\over x}$ is not considered leading,
in that factors of $\alpha_s$ may or may not be accompanied by
$\ln {1\over x}$. Thus, if the LO in the expansion eq.~\laxexpad\ of the
anomalous dimension is used, all logs of the form
\eqn\lls
{\alpha_s^p (\ln Q^2)^q\left(\ln{1\over x}\right)^r }
with $q=p$, and $0\le r\le p$ are summed up.
At NLO the anomalous dimension includes both linear and quadratic
terms in $\alpha_s$, and thus all logs eq.~\lls\ with $q\le p\le 2q$ are
summed (there are at most twice as many powers of $\alpha_s$
as powers of $\ln Q^2$). In standard perturbative computations the
NLO solution is however then linearized, by expanding the exponential
of the NLO term of the anomalous dimensions in powers of $\alpha_s$ and
retaining only the leading nontrivial term (for instance, the NLO
asymptotic correction eq.~\eFasymptwo\ was derived in
this way). This means that only terms with $q\le p\le q+1$, i.e. $p=q+1$
are included at NLO.
Furthermore, there is  an extra power of $\ln {1\over x}$
in the NLO anomalous dimension, so $0\le r\le p$ as at LO.
At NNLO
yet an extra power of $\alpha_s$ per power of $\ln Q^2$ is allowed, and so
forth. We will refer to this as the ``large $x$''
expansion.

This is not the only way to organize the perturbative expansion, however.
We might instead want for instance to consider $\ln {1\over x}$ as leading:
this would be appropriate at very small $x$. Then, all terms
where each extra power of $\alpha_s$ is accompanied by a power of
$\ln {1\over x}$ should be included at LO. These are the leading
singularities
to all orders in $\alpha_s$ (see fig.~4b): it is thus convenient to
reorganize the expansion of the anomalous dimension as
\eqn\smxexpad
{{\alpha_s\over 2\pi}\gamma(N,\alpha)= \sum_{m=1}^\infty \alpha^{m-1}
\Big(\sum_{n=2-m}^\infty A^{n+m-1}_n \bigg(\frac{\alpha}{N}\bigg)^n\Big) ,}
which corresponds to the splitting function
\eqn\smxexpsf
{\eqalign{{\alpha_s\over 2\pi}P(x,t)&=
\sum_{m=1}^\infty \big(\alpha_s(t)\big)^{m-1}
\bigg(\frac{1}{x}\sum_{n=1}^\infty A^{n+m-1}_n \frac{\big(\alpha_s(t)\big)^n
\ln^{n-1}\smallfrac{1}{x}}{(n-1)!}\cr
&\qquad+ \sum_{q=0}^{m-2} A^{m-q-1}_{-q} \big(\alpha_s(t)\big)^{-q}
\frac{d^q}{d\xi^q}\delta(1-x) \bigg).\cr}}
Subsequent orders are still labelled by the index $m$ of the outer sum.
The LO is the sum of leading singularities, and sums all logs
eq.~\lls\ with $r=p$ and $1\le q\le p$.
(terms with $q=0$
are not included because at least one power of $\ln Q^2$ is
produced by integration of the renormalization group equation).
In NLO terms with an extra overall factor of $\alpha_s$ are included
in the anomalous dimension,
so the solution contains terms  with $r< p\le r+q$. Upon linearization,
only terms with $p=r+1$ are kept, while still $1\le q\le p$.
There is then complete symmetry between this ``small $x$'' expansion
and the large $x$ expansion, with the roles of the two
logs interchanged.

The small $x$ approximation to LO evolution discussed in the previous section
corresponds to only retaining the most singular terms in the LO
anomalous dimension: it thus corresponds to taking the intersection of the
small $x$ and large $x$ leading order terms, i.e. the
pivotal term with $m=n=1$ in eq.~\laxexpsf\ or \smxexpsf, which
sums all logs with $p=q=r$. The symmetry of double scaling reflects
this double logarithmic approximation. The terms contributing
to the coefficient $\delta$ eq.~\dasparms\ are large $x$ corrections
(i.e corresponding to
$r<p=q$) and so forth. The fact that double scaling is observed indicates that
the HERA data are taken in a region where the two logs start being
equally important.

This suggests that in this region the most
convenient way of organizing the perturbative expansion is one where
the two logs are treated symmetrically at each order (``double leading''
expansion).
To do this, the anomalous dimensions are expanded as (see fig.~4c)
\eqn\dlexpad
{{\alpha_s\over 2\pi}\gamma(N,\alpha)= \sum_{m=1}^\infty \alpha^{m-1}
\bigg(\sum_{n=1}^\infty A^{n+m-1}_n \bigg(\frac{\alpha}{N}\bigg)^n
+ \alpha \bar\gamma_N^{(m)} \bigg).}
In LO (i.e. when $m=1$) all terms with
$1\le q\le p$,
$0\le r\le p$, and $1\le p\le q+r$ are summed. If all cross terms
(i.e. those containing a product of a contribution to $\bar\gamma_N^{(m)}$
times a singular contribution) are linearized then the solution
includes all contributions where each extra power of
$\alpha_s$ is accompanied by a log of either
$1\over x$ or $Q^2$ or both.\nfig\orders{The terms summed in
the various expansions of the anomalous dimensions and associated
splitting functions: a) the standard (large-$x$) expansion \laxexpad, b)
the small-$x$ expansion \smxexpad, and c) the `double-leading'
expansion \dlexpad. Leading, sub-leading and sub-sub-leading terms are
indicated by the solid, dashed and dotted lines respectively; $m$
denotes the order in $\alpha$, while $n$ denotes the order in $1/N$.
Singular terms are marked as crosses, while
terms whose coefficients known at present (for $\gamma^{gg}_N$)
are marked by circles: the term which leads to double scaling is
marked with a star.} In NLO an overall extra power of $\alpha_s$
will be allowed, and so forth.

Of course, a variety of other expansions which interpolate between
these could be constructed. The crucial point here however is that
all these expansions are consistent with the renormalization group:
in each case  we may define
\eqn\genexp{ {\alpha_s\over 2\pi}\gamma(N,\alpha)=
\sum_{m=1}^\infty \alpha^m \gamma_m(N,\alpha);}
each term in this expansion is then of order $\alpha_s$
compared to the previous one and, in particular, a change of
scale at $k$-th order may be compensated by adjusting the $k+1$-th
order terms. The expansion eq.~\genexp\
may thus be treated using the standard machinery used to
perform NLO or higher order QCD computations.
However, if we choose an expansion which is appropriate
at small $x$ (say, the extreme small $x$ one \smxexpad), then
$\gamma_{m+1}(N,\alpha)$ is of the same order as $\gamma_m(N,\alpha)$
since they both sum up the relevant logarithms, hence the $m+1$-th
order contribution to $\gamma$ is genuinely of order $\alpha_s$ as compared
to the $m$-th order one.
This is not achieved  by an {\it ad hoc} ``resummation''
of a particular
class of contributions, but simply by organizing the perturbative
renormalization group in a different, equally consistent way.

There is a very important issue which must still be addressed however:
namely, in the large $x$ expansion eq.~\laxexpad\
the series in  $n$ which defines
the anomalous dimension at each perturbative order (in $m$) of course
converges --- in fact, we defined this series by expanding out an
expression given as a function of $N$. The summation of leading
singularities eq.~\bfklexp, however, does not converge for all $N$:
as mentioned earlier, $\gamma_{L}(N,\alpha_s)$ has a branch-point,
implying that the series has
a finite radius of convergence, and diverges when $N<4\ln \bar\alpha$
(the other two branch points on the first sheet~\ehw\ being also
inside this circle).
In fact it is not even (Borel) resummable for $\Re N<4\ln \bar\alpha$,
since the integral over the Borel transformed series (which has
infinite radius of convergence) diverges at the upper limit.
This seems to pose an insurmountable problem
for the perturbative approach to small $x$ evolution: the
series which defines the leading coefficient in expansion \smxexpad\
of the anomalous dimension, which was supposed to be useful for small
$N$, is actually only well defined when $N$ is sufficiently large.
This apparent inconsistency seems to have led many to the conclusion that
conventional perturbative evolution breaks down at small $x$.

The dilemma is resolved by the observation that
the physically relevant quantity, namely the splitting function
$P_{gg}(x)$, is instead given at leading order in the expansion
\smxexpsf\ by the series
\eqn\ggspf
{{\alpha_s\over2\pi}
P_{gg}(x,t)={\alpha_s(t)\over x}\sum_{n=1}^\infty (A_n^n)^{gg}
{\left(\alpha_s(t)\ln{1\over x}\right)^{n-1}\over (n-1)!}
\equiv\frac{\lambda_s(t)}{x}B\big(\lambda_s(t)\ln\smallfrac{1}{x}\big),}
where $\lambda(\alpha_s)$
is given by eq.~\lipex\eqnn\convser\eqnn\lipco
$$\eqalignno{
&B(u) =\sum_{n=1}^\infty \frac{a_n}{(n-1)!} u^{n-1}&\convser\cr
&a_n= (A_n^n)^{gg}(4\ln 2\, C_A/\pi)^{-n}.&\lipco}$$
Because the  series $\sum_{n=1}^\infty a_n v^n$
has radius of convergence
one, the radius of convergence of $B(u)$ is infinite, and thus
the series which defines the splitting function eq.~\ggspf\
converges uniformly on any finite intervals of $x$ and $t$ which
exclude $x=0$: the only reason for the bad
behaviour of the series \bfklexp\ is that when transforming to Mellin
space one attempts to integrate all the way down to $x=0$, and this is
not possible for singlet distributions because the total number of partons
diverges there. If instead the parton distributions are evolved using the
Altarelli-Parisi equations \aps, the splitting functions are only
required over the physically accessible region $x>x_{\rm min}$, and
no convergence problems arise. Indeed, because the series for the
splitting function is convergent it is only necessary, when working
in a physically accessible region to a certain level of accuracy, to
retain a finite number of terms. Since then the only singularities in
the Lipatov anomalous dimension are poles at the origin, it follows
that the cuts which arise to all orders are strictly unphysical.

Similar considerations will apply to the subsequent orders in the
expansion of the anomalous dimension eq.~\smxexpad: if the coefficients
in the expansion of the anomalous dimension have a nonvanishing radius of
convergence, then the corresponding splitting functions will
converge for all $x>0$. This will be true provided there is no singularity
(or accumulation of singularities) in the anomalous dimensions
at $N=\infty$ or, equivalently, at $\alpha_s=0$. This is a standard
assumption in perturbative QCD (at least order by order in perturbation
theory), so it is natural to conjecture that this will be true
for all values of $m$ in eq.~\smxexpad.

So far we considered the behaviour of the $\gamma_{gg}$ anomalous dimension.
We already know that at LO only $\gamma_{gg}$ and $\gamma_{gq}$ display
a singularity at $N=0$. In fact, it may be proven~\fact\ that the leading
singularities in $\gamma_{gg}$ and $\gamma_{gq}$ are related to
all orders, according to the so-called color-charge relation
\eqn\colch{\gamma_{gq}={C_F\over C_A}\gamma_{gg}+O\left[\alpha_s\left({
\alpha_s\over N}\right)^n\right],}
while the coefficients of the leading singularities in the quark anomalous
dimension $\gamma_{qq}^{\rm S}$ and $\gamma_{qg}$ vanish to all orders.
The coefficients of the NL singularities (i.e. the NLO in the small $x$
expansion eq.~\smxexpad) have been calculated
recently~\ref\cathau{S.~Catani and F.~Hautmann, \PL\vyp{B315}{1993}{157};
\NP\vyp{B427}{1994}{475}.}: beyond the (nonsingular)
 lowest order in $\alpha_s$ [given in eq.~\smallnad] they also
satisfy a color charge relation, namely
\eqn\colchq{\gamma_{qq}^{\rm S}={C_F\over C_A}\left(\gamma_{qg}-
{4\over 3}T n_f \right)
+O\left[\alpha_s\left({
\alpha_s\over N}\right)^n\right].}
The series expansions for these NLO anomalous dimensions have the same
radius of convergence as that of the Lipatov anomalous dimension,
though the form of the branch point singularity is now scheme dependent.
The NLO coefficients in the expansion of the gluon anomalous
dimensions are still unknown.

All the NLO coefficients, being subleading, are scheme dependent: given their
expression in the parton scheme, where coefficient functions satisfy
eq.~\cf, their expressions in any
other factorization scheme such as the \MS\ scheme
may be determined, along with
the corresponding coefficient functions.
In fact, the freedom of choosing a factorization scheme
turns out~\ref\mom{R.D.~Ball
and S.~Forte, \PL\vyp{B359}{1995}{362}.}
to be wider in all expansions other than the standard large $x$
expansion eq.~\laxexpad: this is due to the fact that the normalization
of the gluon beyond order $\alpha_s$ may be modified by  change of scheme.
It follows that, besides the usual freedom to perform a scheme change
which modifies the $F_2$ coefficient functions, there is now also
the possibility of performing a scheme change which does not affect
the $F_2$ coefficient functions, but changes
the definition of the gluon distribution (while leaving all
LO anomalous dimensions unaffected, as a scheme change ought to).
Such a redefinition, in particular,
affects the quark anomalous dimensions,
which only start at NLO in the small $x$ expansion:
for instance, it can be
used~\ref\ciakp{M.~Ciafaloni, \PL\vyp{B356}{1995}{74}.} to remove a
singularity which
this anomalous dimension has at NLO in the parton scheme~\cathau\
at the Lipatov point $N=\lambda_L(\alpha_s)$.
The nonsingular anomalous dimensions thus obtained
has been argued~\ciakp\ to have a more direct physical interpretation
in that they correspond to specifying the initial parton distributions
in accordance with Regge theory. It is also possible to find a scheme
in which the quark anomalous dimensions vanish~\ref\sdis{S.~Catani,
DFF 226/5/95, {\tt hep-ph/9506357}.}.

This freedom, however, could in principle be completely pinned down by
requiring
that the evolution equations conserve momentum, which implies the
constraints
\eqn\momcon{\gamma_{qg}(1,\as)+\gamma_{gg}(1,\as)=0,\qquad
\gamma_{qq}^{\rm S}(1,\as)+\gamma_{gq}(1,\as)=0,}
in analogy to what happens at large $x$, where the gluon normalization
(which in that case is just a constant)
is fixed by the same requirement. These constraints,
which are necessary consistency conditions that
must be satisfied order by order in the expansion of anomalous dimensions,
cannot be satisfied
at LO in the small $x$ expansion, where however they do not
apply because the the gluon decouples from $F_2$ (which therefore does
not evolve). At NLO they fix uniquely
the NLO coefficients in the small $x$ expansion eq.~\smxexpad\
of $\gamma_{gg}$
in terms of the LO coefficients of $\gamma_{gg}$
and the NLO coefficients of $\gamma_{qg}$~\alphas:
substituting eq.~\smxexpad\
in eq.~\momcon\ implies immediately
\eqn\momconst{(A^{n+1}_n)_{qg}+(A^{n+1}_{n})_{gg}+(A^{n+1}_{n+1})_{gg}=0.}
It can be proven~\mom\ that the freedom of choosing the gluon
normalization by a change of scheme is in one-to-one correspondence
with the values of the NLO coefficients in $\gamma_{gg}$: of course,
whatever the choice
of gluon normalization there exists a set of values of $(A^{n+1}_{n})_{gg}$
which satisfies eq.~\momcon; more interestingly,
whatever the
coefficients $(A^{n+1}_{n})_{gg}$ turn out to be
when they will be calculated,
there exist a choice
of normalization that will enforce eq.~\momcon.\foot{Strictly speaking
this is only possible when $n\ge2$ in eq.~\momcon: the scheme-independent
$O(\alpha_s)$ terms violate momentum conservation
in the small $x$ expansion. This violation is of course asymptotically
subleading as $x\to0$, when this expansion is supposed to hold. Exact
momentum conservation can be obtained order by order
in the double leading expansion
eq.~\dlexpad.} At present we have to content
ourselves with parametrizing our ignorance of these coefficients by
the choice of gluon normalization, and estimating the corresponding
uncertainty by varying this normalization.

In the nonsinglet channel the inclusion of the leading singularities
is more problematic.
The singularity at $N=-1$ in the anomalous dimension
$\gamma_{qq}^{\rm NS}$
is known~\lipkir\ to be stronger than that at $N=0$ of the
singlet anomalous dimensions, i.e. to be double-logarithmic:
\eqn\nslsing{{\alpha_s\over2\pi}\gamma_{qq}^{\rm NS}(N,\alpha)=
\sum_{m=1}^\infty \alpha^m
\sum_{n=-\infty}^{2m-1} \tilde A^m_n (N+1)^{-n}.}
This is in agreement with the two loop result \twoloopsing,
and is indeed the generic expected behaviour of anomalous
dimensions in the small $N$ limit. The behaviour of
the nonsinglet distributions which is obtained by including these
singularities
to all orders in the coupling has been computed~\lipkir, and
leads to a power-like growth: $q_{\rm NS}\sim x^{-\lambda_{\rm KL}}$
with $\lambda_{\rm KL}=\sqrt{8\alpha/3\pi}$, with an almost identical
result for charge-conjugation odd distributions. This is in fact very
similar to the growth predicted by Regge theory and discussed in the
previous section. However, it is obtained with fixed coupling and is
not yet consistent with the renormalization group. Since the
corresponding factorization theorems are as yet unproven, it is still
unclear whether it will be possible to treat this case in the same way
as the singlet one, making it consistent with the renormalization
group by using the results
of ref.~\lipkir\ to derive the leading term (namely the coefficients
$\tilde A^m_{2m-1}$~\ref\EMR{B.I.~Ermolaev S.I.~Manayenkov and M.G.~Ryskin,
      DESY 95-017, {\tt hep-ph/9502262}.}) in the small $x$ expansion
of the nonsinglet splitting function which can then be used in
the Altarelli-Parisi equation in the same way as in the singlet
sector. What is clear is that the effect of such a
procedure would be relatively undramatic, given that the boundary
condition is already hard, with $\tilde\lambda\sim\lambda_{\rm KL}$.
It follows that the singlet contribution to $F_2^p$ still dominates
the nonsinglet at small $x$, and thus for the rest of the discussion
we ignore it and consider only the singlet.

We can now finally study the small $x$ behaviour of singlet parton
distributions: using evolution equations which include logarithmic effects
in $\ln{1\over x}$ to all orders we will be able to assess in which region
the simple LO double scaling predictions are reproduced.
Even though in the HERA region the double leading expansion is probably
more appropriate, let us consider the extreme small $x$ expansion
eq.~\smxexpad, both because it is somewhat simpler, and because
it allows us to test the stability of double scaling in an extreme case:
if the LO correction in this expansion does not spoil scaling,
no other corrections will since all the most singular corrections are
included, and they all have the same sign.

In the small $x$ expansion at LO the quark anomalous dimensions
vanish, hence the quark sector may be neglected, and the
gluon evolution equation is found using
the splitting function $P_{gg}$ eq.~\smxexpsf\ in the Altarelli--Parisi
equation
eq.~\aps:
\eqn\gapeq
{\eqalign{{\partial \over \partial \zeta} G(\xi,\zeta)
&= {4C_A\over \beta_0}\sum_{n=1}^\infty
a_n\frac{\lambda_s(\zeta)^{n-1}}{(n-1)!}
\int_{-\xi_0}^\xi \!d\xi^\prime\,(\xi-\xi^\prime)^{n-1}
G(\xi^\prime,\zeta)\cr
&= \gamma^2\sum_{n=0}^\infty a_{n+1}\lambda_s(\zeta)^{n}
\int_{-\xi_0}^\xi \!d\xi_1\,\int_{-\xi_0}^{\xi_1} \!d\xi_2\,\dots
\int_{-\xi_0}^{\xi_n} \!d\xi^\prime\,
  G(\xi^\prime,\zeta),\cr}}
where we have used the one loop form of $\alpha_s(t)$
as appropriate for a LO calculation,
the coefficients $a_n$ are given by eq.~\lipco, and $\gamma$ is as in
eq.~\dasparms.
Since we retain only singular contributions to the splitting
functions the lower limit $\xi_0=\ln{1\over x_0}$ in the integrations
on the \rhs\ can be consistently set to zero.
Differentiating both sides w.r. to $\xi$ this can again be cast in the form
of a wave equation
\eqn\aoweq
{\eqalign{{\partial^2\over\partial\xi\partial\zeta} G(\xi,\zeta)&=
\gamma^2 G(\xi,\zeta)+\gamma^2\sum_{n=1}^\infty
a_{n+1}\lambda_s(\zeta)^n
\int_{0}^\xi \!d\xi_1\,\int_{0}^{\xi_1} \!d\xi_2\,\dots
\int_{0}^{\xi_{n-1}} \!d\xi^\prime\,
 G(\xi^\prime,\zeta)\cr
&=\gamma^2 G(\xi,\zeta)+\gamma^2\sum_{n=1}^\infty
a_{n+1}\frac{\lambda_s(\zeta)^n}{(n-1)!}
\int_{0}^\xi \!d\xi^\prime\,(\xi-\xi^\prime)^{n-1}G(\xi^\prime,\zeta),\cr}}
which, when only the first singularity is retained,
reduces to eq.~\weq\ with $\delta=0$ (the term contributing to $\delta$ is
NLO in the small $x$ expansion).

The quark does not evolve at LO, hence, in order
to determine the evolution of $F_2$, we must go to NLO. This can be done
in complete analogy to the derivation of the double scaling evolution
equations eq.~\weq\ in the previous section: in fact the derivation
there can be viewed as a simplified version of a NLO calculation in
the small $x$ scheme, where only the (order $\alpha_s$)
contributions with $n=2-m$
to the anomalous dimensions eq.~\smxexpad\ are retained. Thus,
we first diagonalize the anomalous dimension matrix, which up to NLO
has eigenvalues
\eqn\aoeval
{\eqalign{\lambda_+(\alpha)&= \gamma_0^{gg}
+\alpha
\big(\gamma_{gg}^1+\frac{\gamma_{gq}^0}{\gamma_{gg}^0}\gamma_{qg}^1\big)
+O(\alpha^2)\cr
\lambda_-(\alpha)&=\alpha\big(\gamma_{qq}^1-
\frac{\gamma_{gq}^0}{\gamma_{gg}^0}\gamma_{qg}^1\big)+O(\alpha^2).\cr}}
The corresponding eigenvectors are given by
\eqn\aoevec
{Q_+ = \alpha \frac{\gamma_{qg}^1}{\gamma_{gg}^0}G_+
+O(\alpha^2)\qquad Q_- = - \frac{\gamma_{gg}^0}{\gamma_{gq}^0}G_-
+O(\alpha).}
where the anomalous dimension is expanded as in eq.~\genexp.
Notice that the large eigenvector condition does
not depend on the unknown NLO gluon anomalous dimensions, and the
small eigenvector condition has still the simple
form of eq.~\xevec, thanks to the color-charge relation eq.~\colch.

Transforming to $x$ space, the large eigenvector component $G_+$ is seen
to satisfy eq.~\aoweq, plus $O(\alpha_s)$ corrections, which lead to
the $\delta$ term of eq.~\weq\ and to $O(\alpha_s)$ corrections to
the coefficients $a_n$ in eq.~\aoweq.
The quark equation is then found differentiating the large eigenvector
condition wih respect to $\zeta$ and using the equation for $G_+$,
with the result
\eqn\qg
{\frac{\partial}{\partial\zeta}Q_+(\xi,\zeta)
=\frac{n_f}{9}\gamma^2 \Big[ G_+(\xi,\zeta)
+\sum_{n=1}^\infty \tilde a_n \lambda_s(\zeta)^n\int_0^\xi\! d\xi'\,
\frac{(\xi-\xi')^{n-1}}{(n-1)!} G_+(\xi',\zeta)\Big],}
where the coefficients $\tilde a_n$ re given by
\eqn\qlipco{\tilde a_n= (A_n^n)^{qg}(4\ln 2\, C_A/\pi)^{-n}.}

Although these evolution equations cannot be solved in closed form,
a solution can be developed perturbatively in the usual way by noting,
as we did at two loops \eGtwo, that the solution \eGoursat\ for $G_+$
now acquires an extra set of terms
\eqn\eGal{
\gamma^2\sum_{n=1}^\infty
\frac{a_{n+1}}{(n-1)!}
\int_0^\xi\int_0^{\xi'} \int_0^\zeta\!d\xi'd\xi''d\zeta'\,
I_0\big(2\gamma\sqrt{(\xi-\xi')(\zeta-\zeta')}\big)
\lambda_s(\zeta')^n(\xi'-\xi'')^{n-1}G_+(\xi'',\zeta')}
on the \rhs. The full solution may then be developed iteratively by
substitution of lower order solutions, leading to a power series
expansion in terms of integrals over Bessel functions, which can
then be evaluated
numerically. A similar expansion may be obtained in Mellin space by
integrating explicitly the singlet renormalization group
equation \rgeq, and then performing the inverse transform \apesol\ by
choosing a contour which encircles the singularities in the region
$\vert N\vert <\lambda_L$~\ref\furbi{J.~R.~Forshaw,
R.~G.~Roberts and R.~S.~Thorne, \PL\vyp{B356}{1995}{79}.}.

For present purposes, however, we derive some simple analytic
estimates which demonstrate analytically the nature of the solution~\summ.
The key observation which dictates the structure of the solution is
the simple property of Bessel function $z^{n}I_{n-1}(z)=
\smallfrac{d}{dz}\big(z^nI_n(z)\big)$. This implies that if we determine
iteratively the solution to the full gluon equation eq.~\aoweq\ by
then  eq.~\aoweq\ with the perturbation eq.~\eGal\
takes at lowest order of the iteration the approximate form
\eqn\aoweqnapprox{
{\partial^2\over\partial\xi\partial\zeta} G(\xi,\zeta)\simeq
\gamma^2 G(\xi,\zeta)+{2\over \beta_0}\sum_{n=1}^\infty
a_{n+1}\bigg(\frac{\rho\lambda_s(\zeta)}{\gamma}\bigg)^n I_n(2\gamma\sigma).
}
But all Bessel functions in this expansion depend only on the
scaling variable $\sigma$ and have the same asymptotic behaviour eq.~\ebess\
(which only sets in more slowly for higher values of $n$).
Since $\rho^n I_n(2\gamma\sigma)$ is bounded above by $\xi^n
I_0(2\gamma\sigma)$ it follows that double scaling will always set in
asymptotically provided the series $\sum_{n=1}^\infty a_{n+1}
\big(\smallfrac{\xi\lambda_s(\zeta)}{\gamma}\big)^n$ converges
uniformly, that is provided $\xi<\frac{\gamma}{\lambda_s(\zeta)}$.
This is much wider region than that in which double scaling would hold if we
were to impose a ``hard'' boundary condition of the form $x^{-\lambda}$
at $\zeta=0$, namely $\xi<\frac{\gamma}{\lambda_s(0)}$, because
$\lambda(\zeta)$
rapidly falls as $\zeta$ increases. Such a boundary condition is precisely
that which the simple argument based on the location of the singularity in
$\gamma_{L}$ eq.~\bfklexp\ would suggest.

To see more clearly how such a power--like behaviour could arise,
consider the evolution equation eq.~\weq\
in the region of extremely small $x$, where the higher order terms give
the dominant contribution to the
series on the right hand side.
Because the series \convser\ has unit radius of convergence
it follows that $\neath{\rm lim}{n\to\infty}
\frac{a_{n+1}}{a_n}\to 1$. But then, setting $a_{n+1}\approx a_n$ in the sum
in eq.~\aoweq, shifting the summation index by one unit, and using
eq.~\gapeq, we have
\eqn\recur
{{2\over \beta_0}\sum_{n=1}^\infty
a_{n+1}\lambda_s(\zeta)^n
\int_{0}^\xi \!d\xi_1\,\int_{0}^{\xi_1} \!d\xi_2\,\dots
\int_{0}^{\xi_{n-1}} \!d\xi^\prime\,
 G(\xi^\prime,\zeta)=\lambda_s(\zeta) {\partial G\over \partial \zeta}.}
Hence, asymptotically as $\xi\to\infty$ the evolution equation \aoweq\
becomes simply
\eqn\damwe
{{\partial^2\over\partial\xi\partial\zeta} G(\xi,\zeta)
- \lambda_s(\zeta){\partial G\over \partial\zeta}=
\gamma^2 G(\xi,\zeta),}
and the summation of all leading singularities
leads to a damping term in the wave equation.

We can now see how the $x^{-\lambda}$ behaviour obtains in the
fixed-coupling limit:
if the coupling is frozen, then
$\lambda$ is just a constant, and the solution to eq.~\damwe\
is given in terms of the (double scaling)
solution $G_0(\xi,\zeta)$ to the original wave equation
\eqn\lipsol
{G(\xi,\zeta)= e^{\lambda\xi} G_0(\xi,\zeta)=x^{-\lambda}G_0(\xi,\zeta).}
However there is no reason to fix the coupling. A solution with running
coupling can be derived by the saddle point method~\summ, and
turns out to give the double scaling behaviour (up to a small correction)
\eqn\saddas
{G(\xi,\zeta)\sim \frac{1}{\sqrt\sigma}
e^{2\gamma\sigma+(\lambda(0)-\lambda(\zeta)\rho^2},}
throughout the region
 $\xi\ll \frac{\gamma^2}{(\lambda(0)-\lambda)^2} \zeta^3$. When $\xi$ is
extremely large the power-like behaviour
\eqn\sadlip
{G(\xi,\zeta)\sim \frac{1}{\xi}
\bigg(\frac{\xi}{\lambda(0)-\lambda(\zeta)}\bigg)^{\gamma^2/\lambda_0}
e^{\xi\lambda(0) + \gamma^2\zeta/\lambda(0)}}
is found instead. This has essentially the ``hard'' form of eq.~\lipsol,
with the large value of $\lambda$ evaluated at the initial scale,
but it is confined to the extremely small region
$\xi\gg\frac{\gamma^2}{\lambda(0)^2}e^\zeta$.
In fact this is still an overestimate of the region where the power-like
behaviour should arise: approximating the evolution equation with eq.~\damwe\
ignores the fact that the asymptotic behaviour of the coefficients $a_n$
only sets in rather slowly, and, by the time it does, $a_n<<a_1$.
If this
effect is taken into account the power-like region is further reduced.

Even though of course all these estimates are only based on the asymptotic
form of the coefficients $a_n$, and thus they will not accurately reproduce
subasymptotic and specifically subexponential corrections (such as the
factor of $1\over\sqrt{\sigma}$ in eq.~\saddas) and normalizations,
they do correctly
estimate the leading behaviours and their region of
validity, as confirmed by a
full numerical analysis.
This allows us to conclude that, even though
a power like behaviour of the form eq.~\lipsol\ is generated very close to
the $\zeta=0$ boundary, it very rapidly dies off due to the running of
the coupling, rather than spreading in the
whole $\xi>{\gamma\over\lambda(0)}$
region as it would do if it were input to LO evolution. Furthermore,
any hard power-like boundary condition will be hidden by this rise
(unless it is even stronger than it): since $\lambda(0)$ is very large
for reasonable choices of the starting scale (for instance
$\lambda(4~{\rm GeV}^2)\approx 0.8$) this means than scaling will
appear very rapidly for all boundary conditions, except unreasonably
hard ones such as $x^{-\lambda}$ with $\lambda\lsim -0.8$ at $Q^2=4$~GeV$^2$.

Turning finally to the quark equation, we may take advantage of the
fact that,
throughout most of the $(\xi,\zeta)$ plane, $\bar G(\xi,\zeta)\simeq
{\cal N}I_0(2\gamma\sigma)$: substituting this in the r.h.s. of eq.~\qg,
and neglecting the small eigenvector we get
\eqn\qgapprox
{\frac{\partial}{\partial\zeta}Q(\xi,\zeta)\simeq\frac{n_f}{9}\gamma^2
{\cal N}\sum_{n=0}^\infty \tilde a_n
\bigg(\frac{\rho\lambda_s(\zeta)}{\gamma}\bigg)^n I_n(2\gamma\sigma).}
It follows that when $\xi\ll\frac{\gamma^2}{\lambda_0^2}\zeta
e^{2\zeta}$, $Q(\xi,\zeta)$ still scales;
for
$\xi\gg\frac{\gamma^2}{\lambda_0^2}\zeta e^{2\zeta}$, if we set
$\tilde a_n = 1$ for all $n$, \qgapprox\ may be simplified yet further
by using the relation $\sum_{-\infty}^{\infty}t^n I_n(z) =
\exp\big(\half z(t+t\inv)\big)$, to give
\eqn\qpower
{Q(\xi,\zeta)\sim {\cal N}\zeta
e^{\xi\lambda_0+\gamma^2\zeta/\lambda_0}.}

The quark anomalous dimensions may thus still produce a growth
of $F_2$ close to the boundary, which however again does not
spread in the $(\xi,\zeta)$ plane due to the running of the coupling.
The actual size of the region where this growth will appear
depends on the size of the coefficients $\tilde a_n$. This, in turn,
strongly depend on the choice of gluon normalization which, as discussed
above, can be modified by changing factorization scheme. The latter could
be fixed using momentum conservation~\mom\ if we knew the NLO small $x$
gluon anomalous dimensions. Since we do not, the effect of the quark
anomalous dimensions on the relative normalization of quark and gluon
distributions (and thus on the relative size of $F_2$ and $F_L$) turns out to
be strongly dependent on the choice of factorization scheme: while the
$\tilde a_n$ corresponding to the \MS\ or DIS schemes~\cathau\
give large (but very different) effects, the $Q_0$-schemes~\ciakp,
being less singular, lead to substantially smaller effects, while in
some schemes~\sdis\ there is almost no effect at all~\ref\paris{S.~Forte
and R.~D.~Ball, CERN-TH/95-184, {\tt hep-ph/9507211}, to be published in
the proceedings of the {\it ``Workshop on Deep Inelastic Scattering
and QCD'' (DIS 95)}, Paris, April 1995.} except at very small $x$ very
close to the boundary.

These analytic results are all closely supported by detailed numerical
investigations, either retaining only the singular terms in the
anomalous dimensions~\refs{\summ,\furbi}, or including also the
full one and two loop contributions in the double leading
scheme~\refs{\alphas,\paris} or using some other procedure~\ehw.
When care is taken to consistently factorize
residual nonperturbative effects into the boundary conditions, all
these analyses reach essentially the same conclusions. The leading (Lipatov)
singularities have only a negligible effect throughout the measured
region, so small that there is still no empirical evidence for them,
not only because they don't affect the shape of $F_2$
in most of the $(\xi,\zeta)$ plane, but also
because the coefficients $a_n$ are so small. The
subleading quark anomalous dimensions have little effect of
the shape of $F_2$, explaining the success of conventional
perturbation theory and double scaling \refs{\summ,\alphas}. However
they can have a substantial effect on the scale $Q_0$ at which soft
initial distributions are input, raising it to around $2\GeV$ in the \MS\
scheme\refs{\summ,\ehw}, and on the relative normalization of quark
and gluon, and thus on the size of $F_L$ as deduced from the measured
$F_2$ \refs{\summ,\furbi}. However these latter effects are
strongly factorization scheme dependent, both in sign and magnitude
\paris, and this scheme dependence could only be resolved
theoretically by a complete calculation of the subleading
singularities of the gluon
anomalous dimension, or phenomenologically by a direct measurement of $F_L$.

\bigskip
\newsec{Outlook}
\medskip
The observation of double asymptotic scaling at HERA is a striking success
of a perturbative QCD prediction made now more than 20 years ago~\DGPTWZ.
The explanation of the effect turns out to be somewhat
subtler than it might seem at first: even though it
is a direct consequence of the singularity structure of the
leading-order perturbative evolution equations, its stability in a
region where higher order effects might naively be expected to be
important is due to the all order cancellation of double logarithmic
singularities, the unexpected smallness of the coefficients of the
remaining single logarithmic singularities, and then finally
the reduction in their impact when the effect of the running
coupling is included in a way consistent with the renormalization group.

This result has interesting ramifications from both the phenomenological and
the theoretical point of view, which have just started to be explored.
The success of NLO perturbation theory in the HERA region strongly
suggests that this may be an ideal place to perform precision tests
of perturbative QCD, since most of the poorly known low-energy effects
(such as higher twists) are either absent or negligible here.
For example, $\alpha_s$ can be measured from existing HERA data~\alphas\ with a
precision comparable to that of all other existing deep-inelastic
experiments combined. As more data in the very small $x$, not-so-small $Q^2$
region become available, they may shed light on subtle issues of scheme
dependence, and provide information on the behaviour of parton
distribution which are input to perturbative evolution.
On the more theoretical side, it would be desirable to derive perturbation
theory at small $x$ in a way which treats the two large scales symmetrically
{}~from the outset, rather than solving renormalization group equations
with respect to one scale while including the summation of the other scale
in the anomalous dimensions, as we did here. This, besides
being interesting for its own sake, could shed
light on the dynamics of perturbative QCD in the high energy regime.

\medskip
\noindent{\bf Acknowledgements:} S.F. thanks Maciej Nowak for organizing
an interesting interdisciplinary school, and K.~Golec-Biernat and
L.~Lipatov for stimulating discussions during the school.

\goodbreak

\medskip
\listrefs
\listfigs
\vsize=27truecm
\null
\vskip -4.truecm
\epsfxsize=14truecm
\hfil\epsfbox{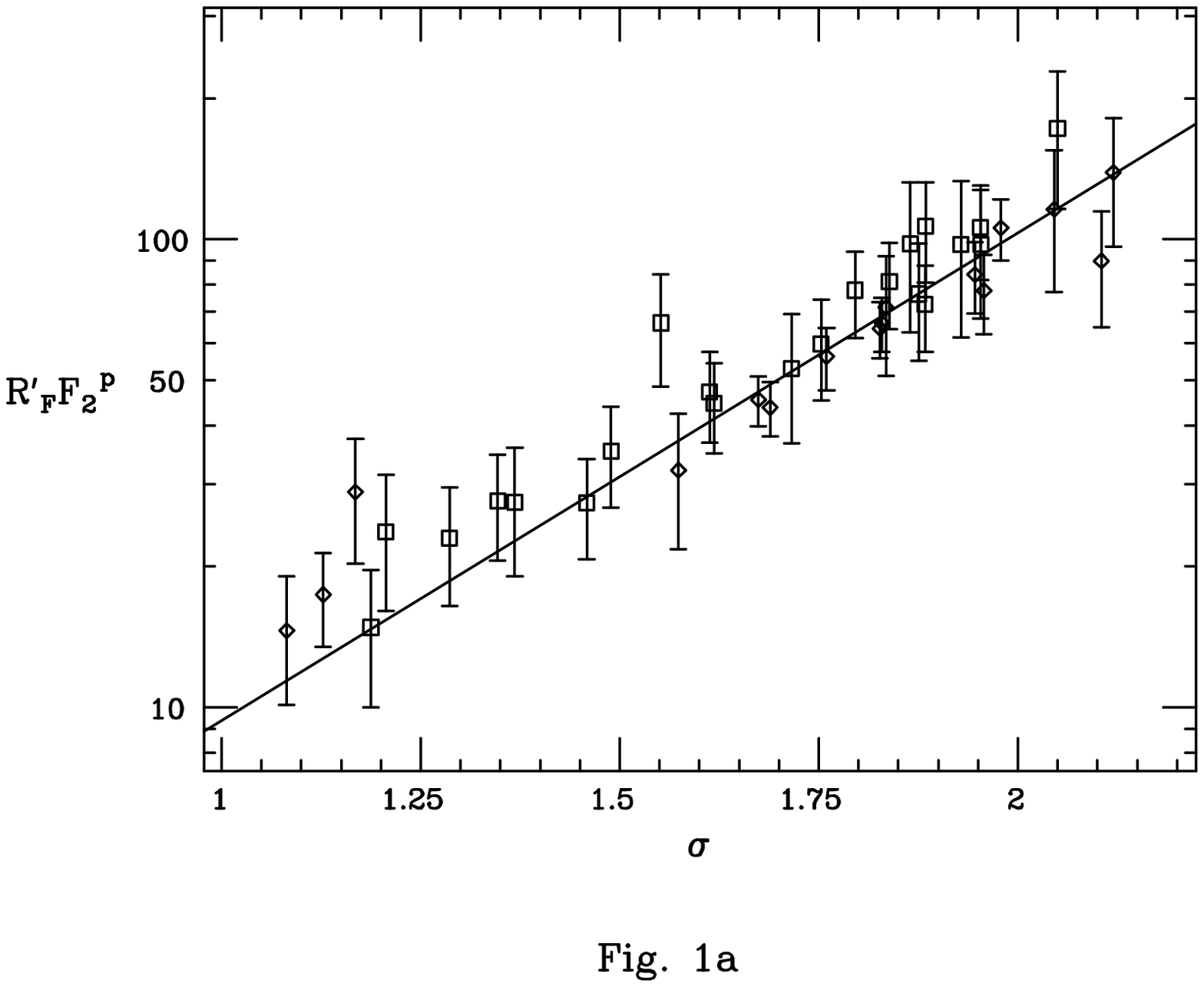}\hfil
\smallskip
\vskip -7.truecm
\epsfxsize=14truecm
\hfil\epsfbox{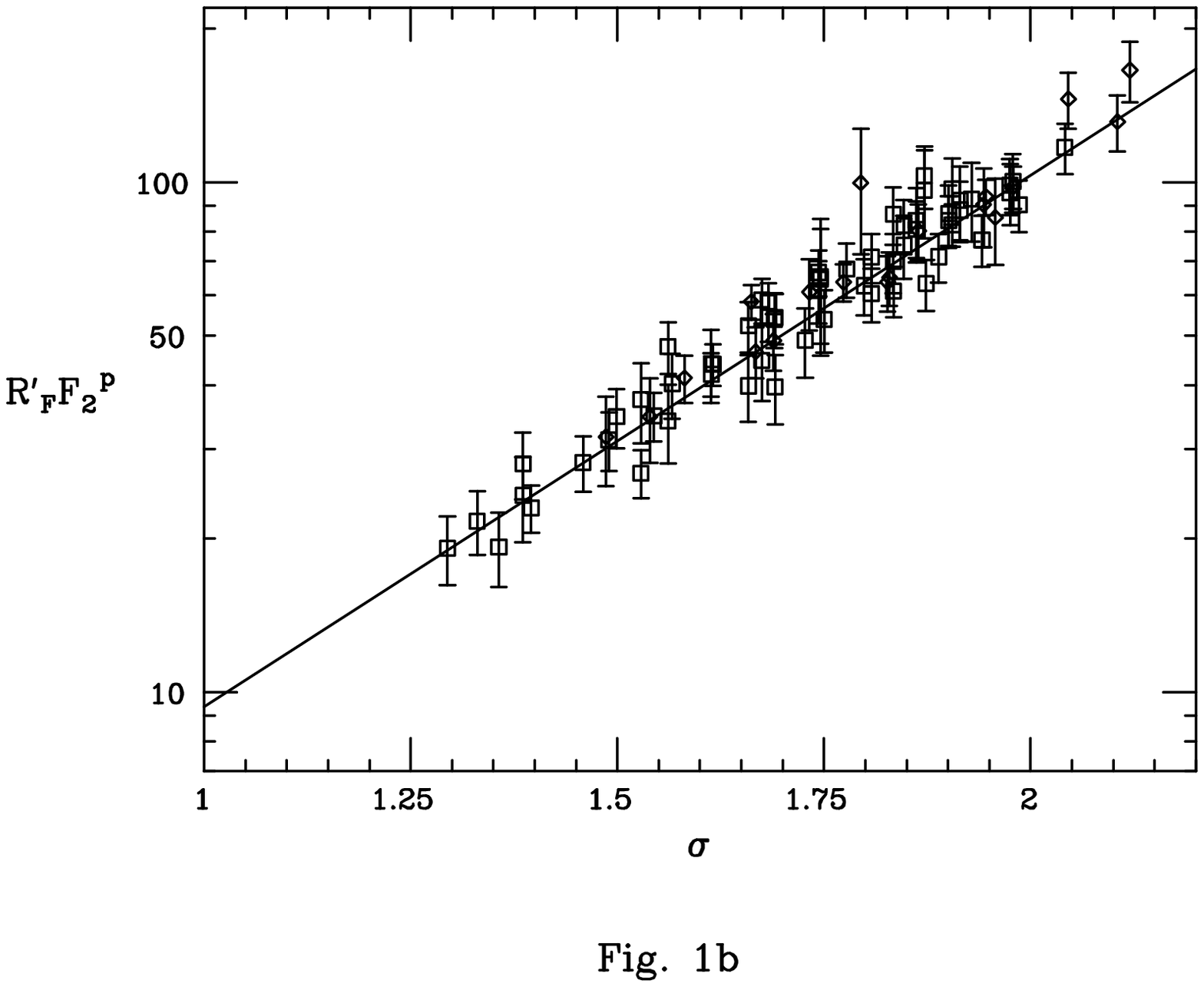}\hfil
\vfill
\eject
\null
\vskip -4.truecm
\epsfxsize=14truecm
\hfil\epsfbox{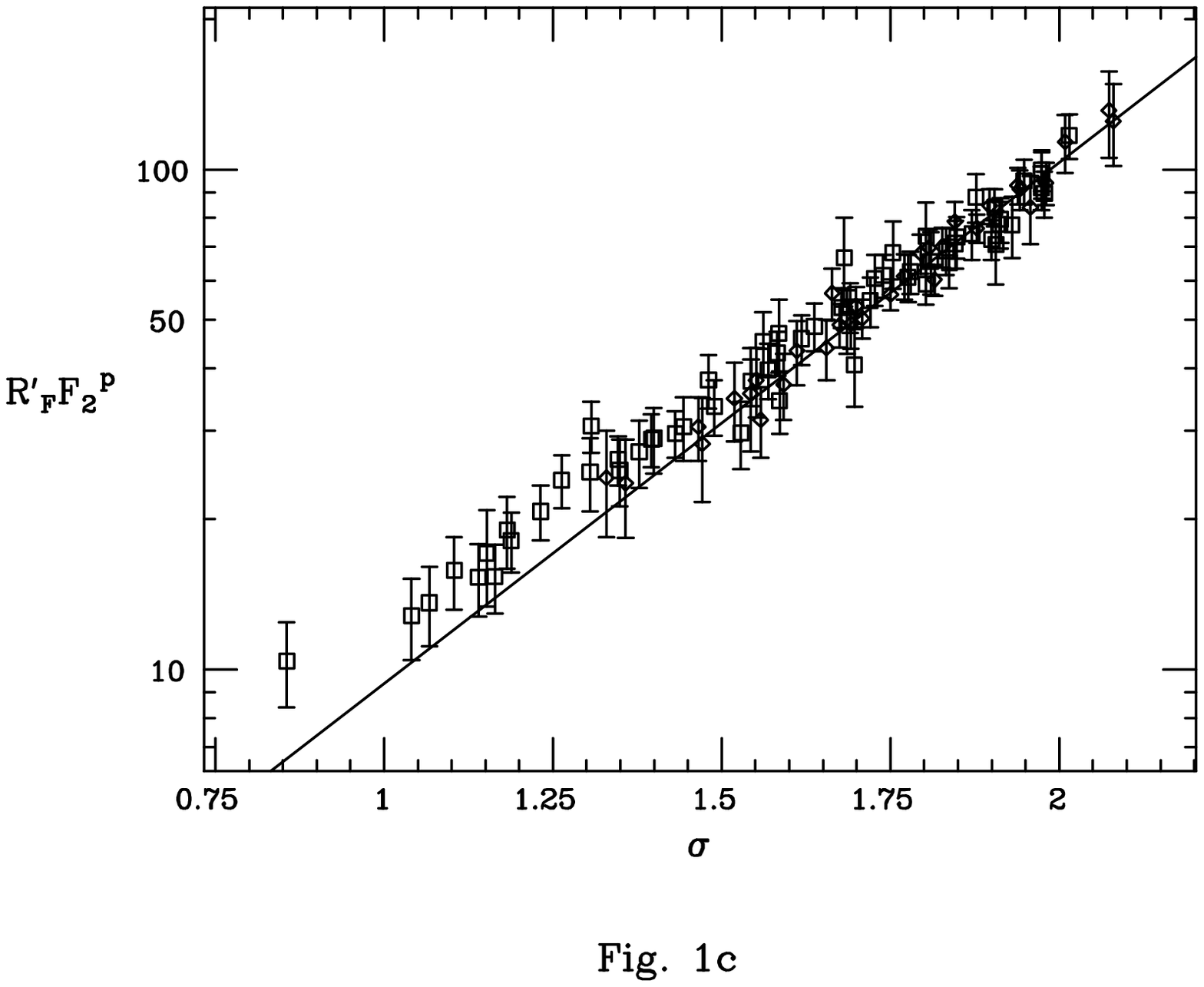}\hfil
\bigskip
\epsfxsize=14truecm
\hfil\epsfbox{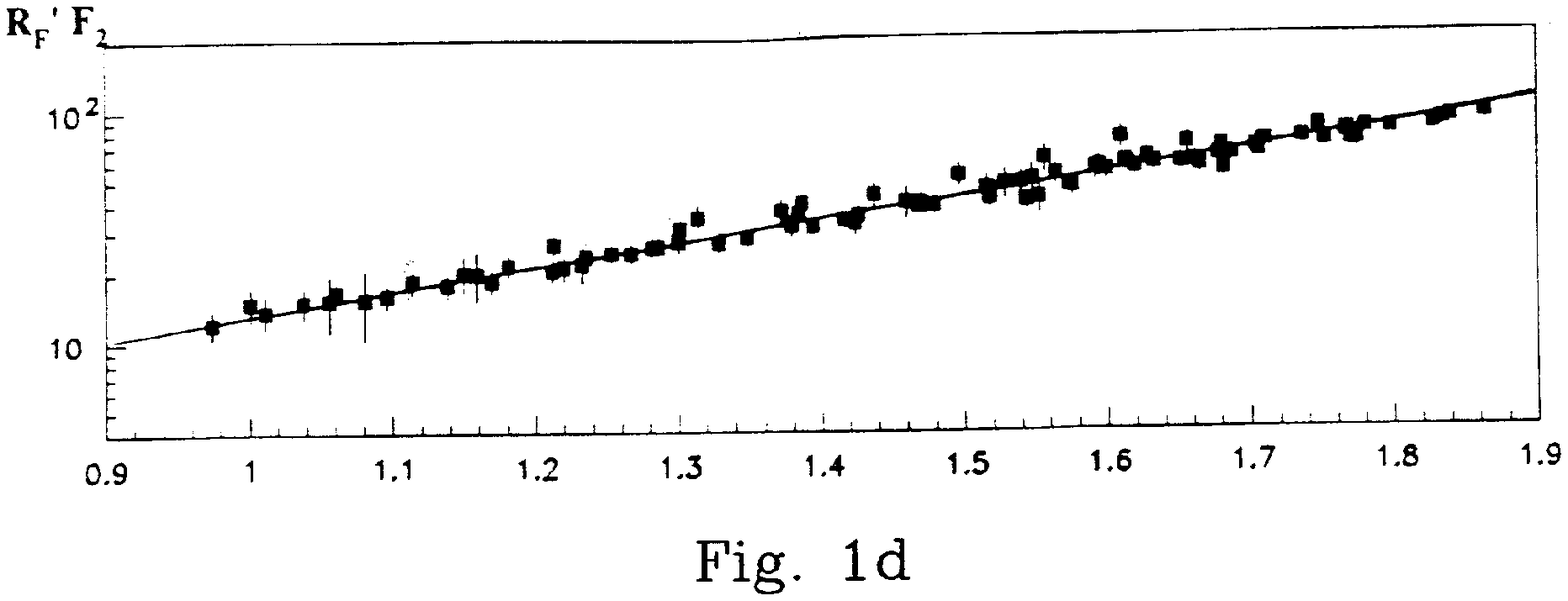}\hfil
\vfill
\eject
\vskip -4.truecm
\epsfxsize=16truecm
\hfil\epsfbox{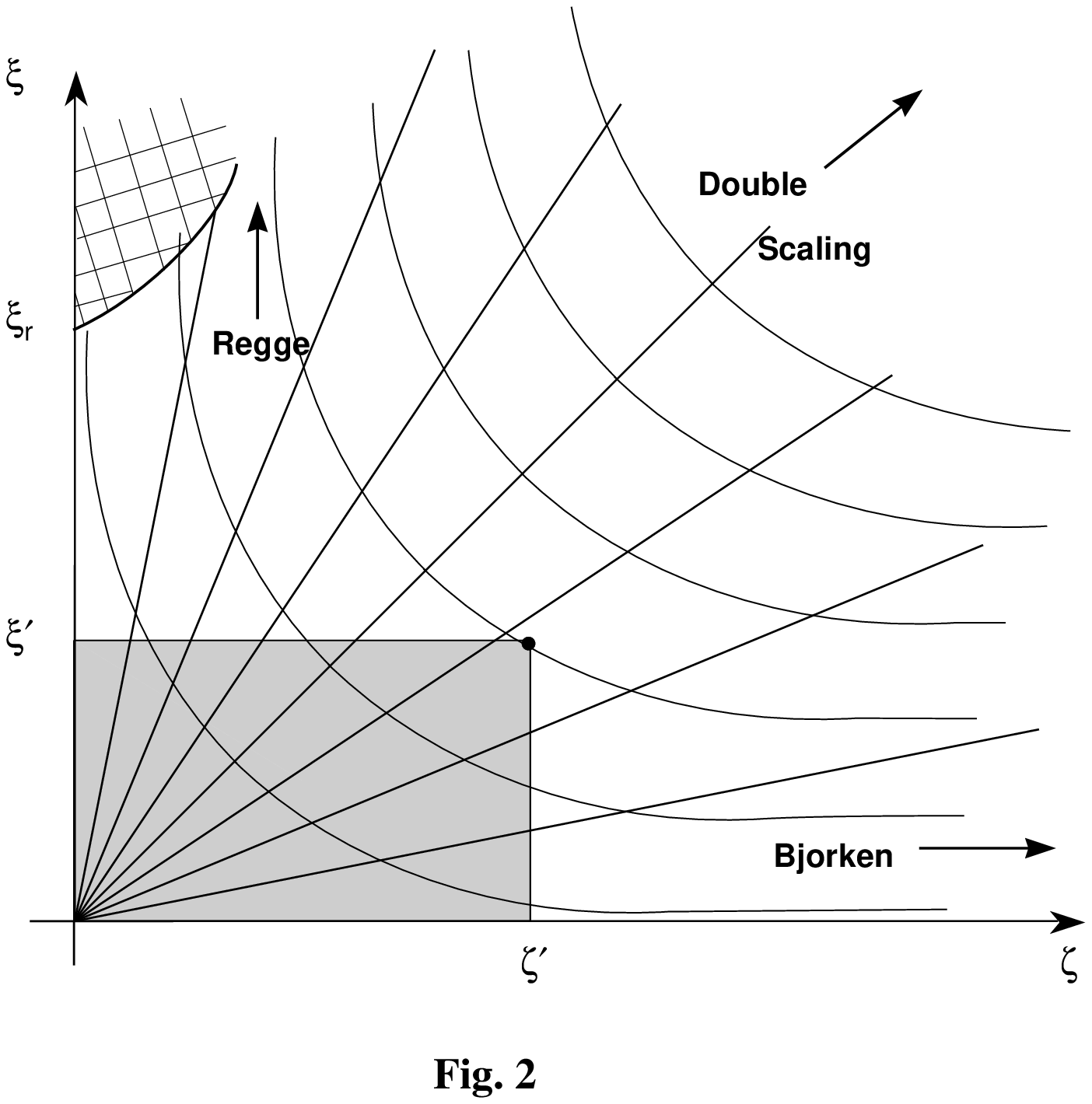}\hfil
\vfill
\eject
\null
\vskip -4.truecm
\epsfxsize=14truecm
\hfil\epsfbox{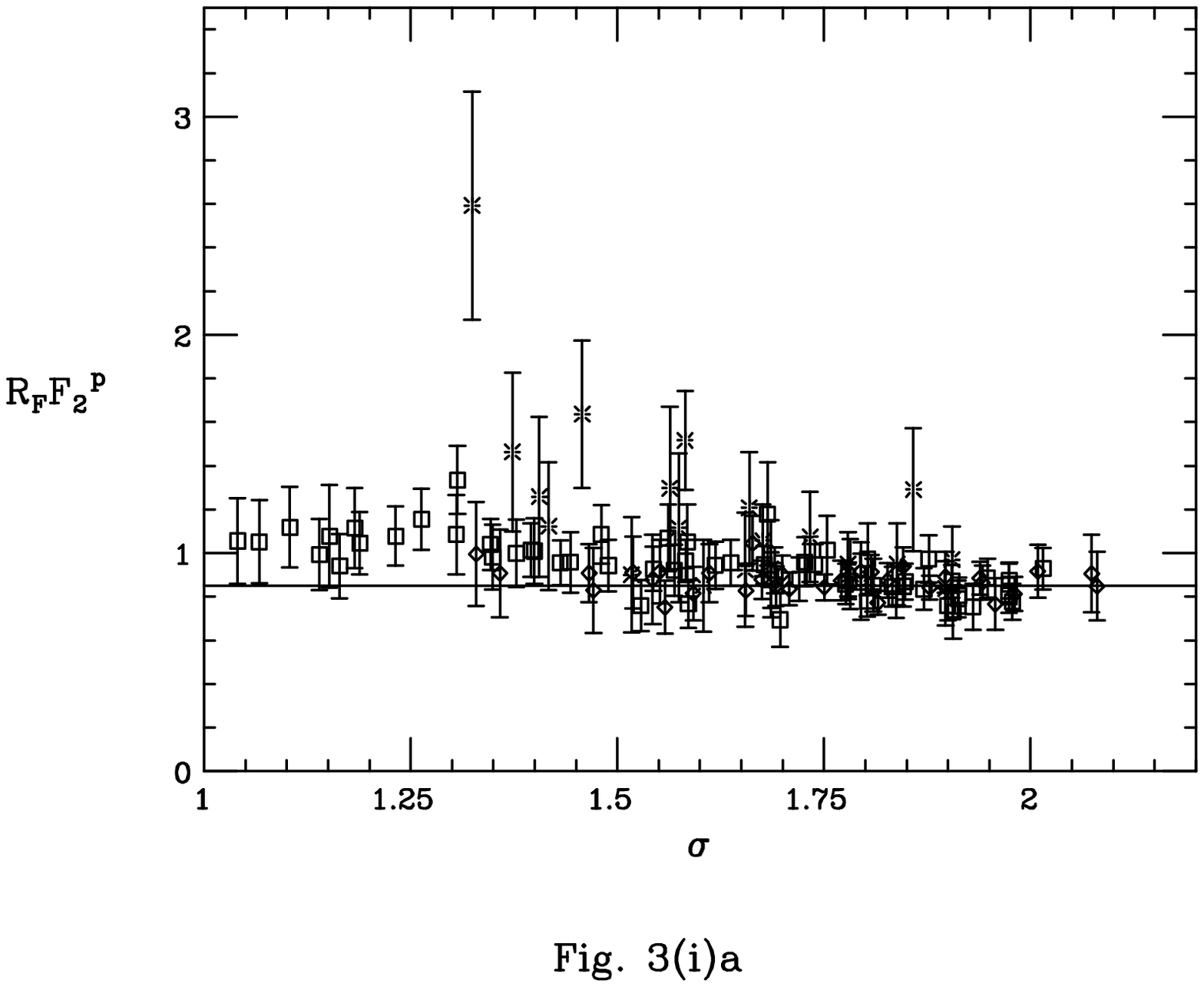}\hfil
\smallskip
\vskip -7.truecm
\epsfxsize=14truecm
\hfil\epsfbox{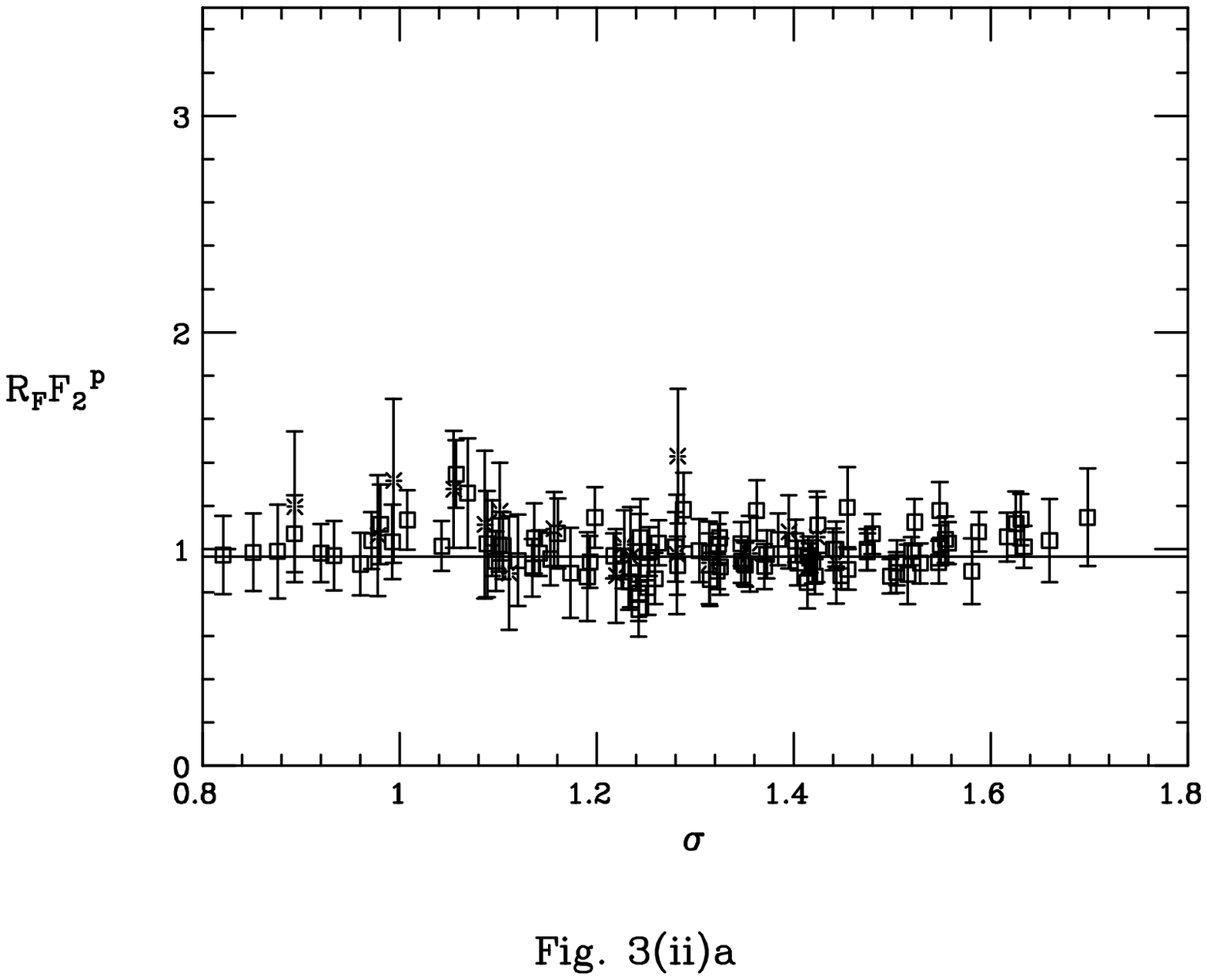}\hfil
\vfill
\eject
\null
\vskip -4.truecm
\epsfxsize=14truecm
\hfil\epsfbox{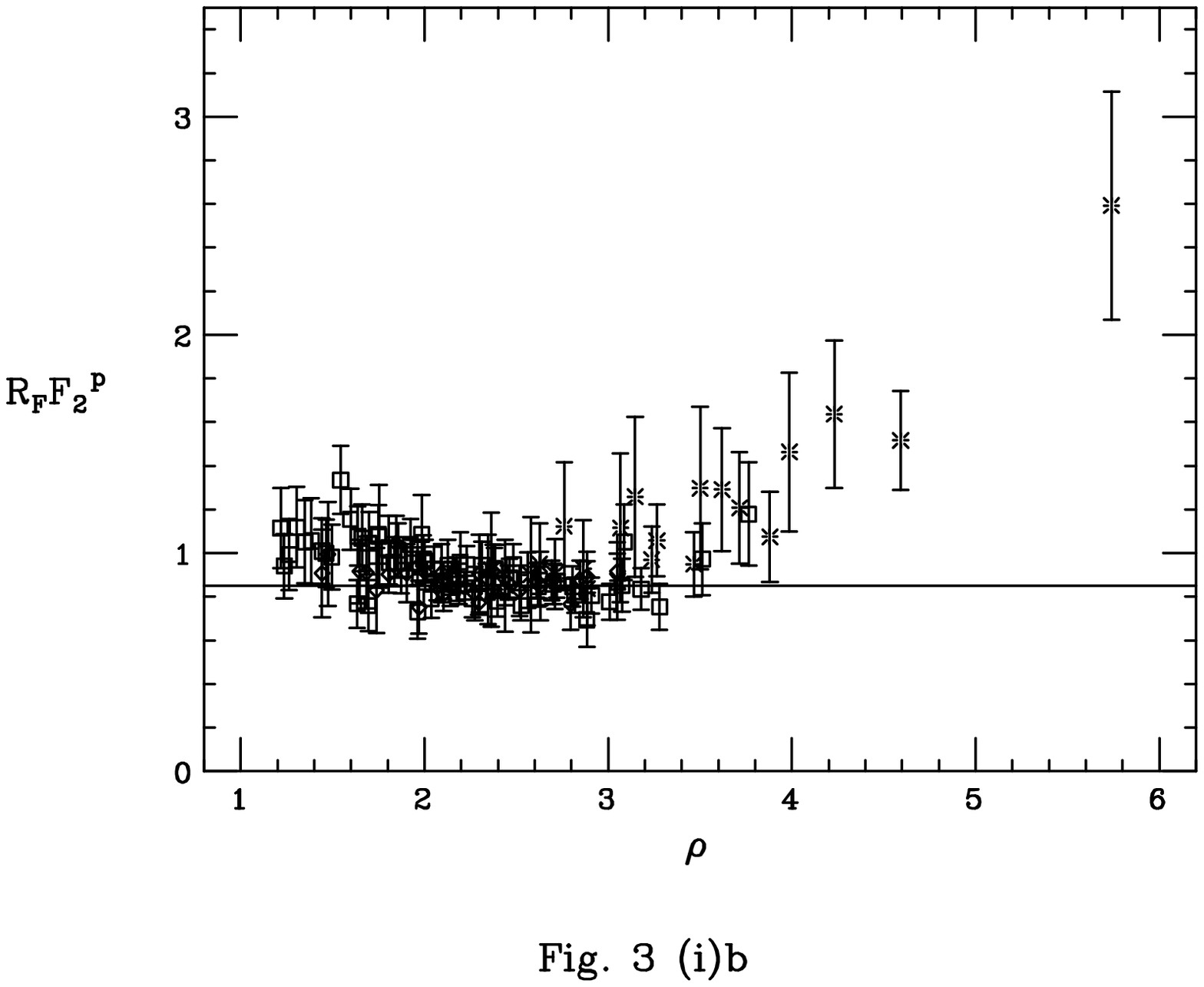}\hfil
\smallskip
\vskip -7.truecm
\epsfxsize=14truecm
\hfil\epsfbox{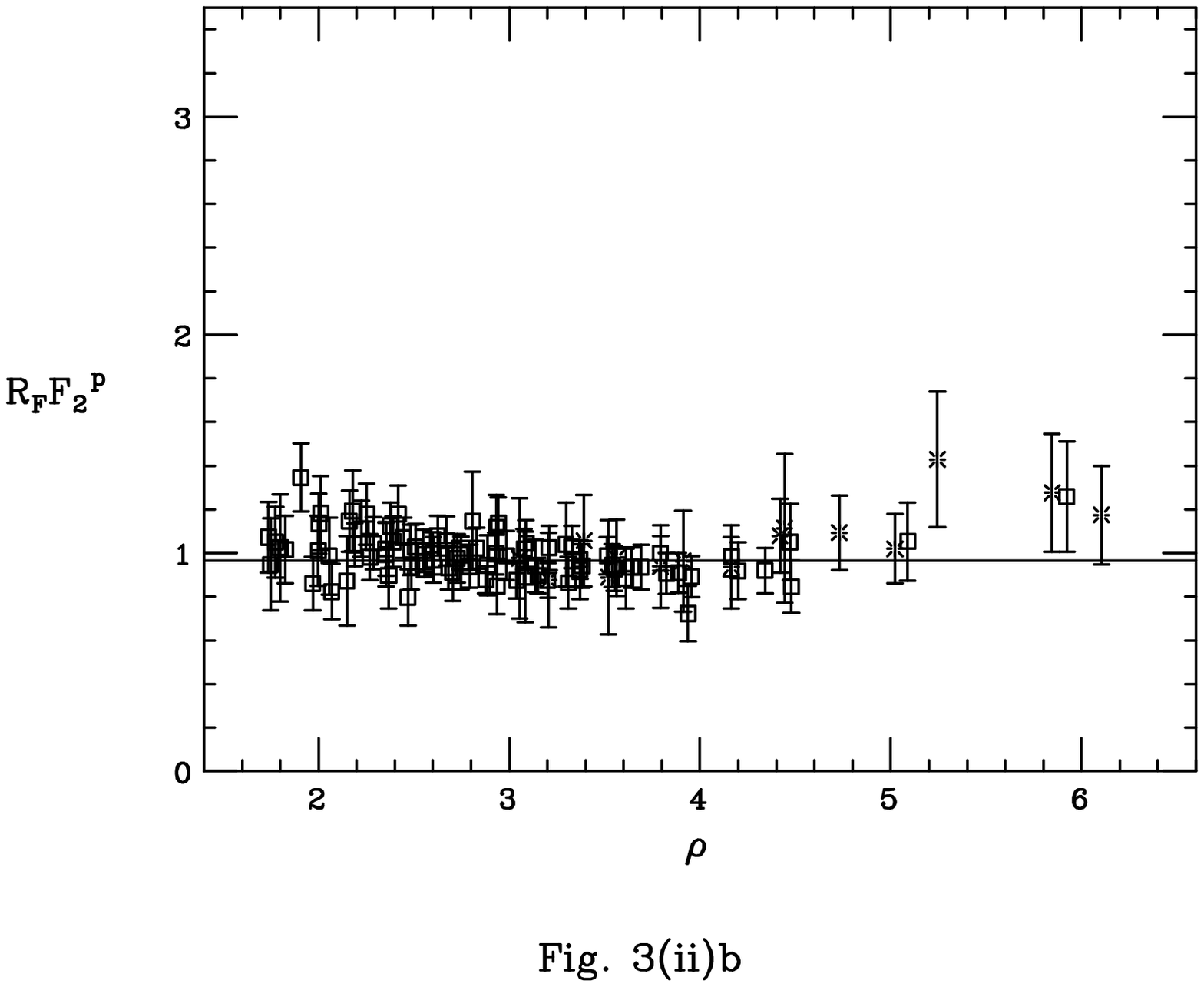}\hfil
\vfill
\eject
\hbox{\hfil\hskip-4truecm
\epsfxsize=10truecm\epsfbox{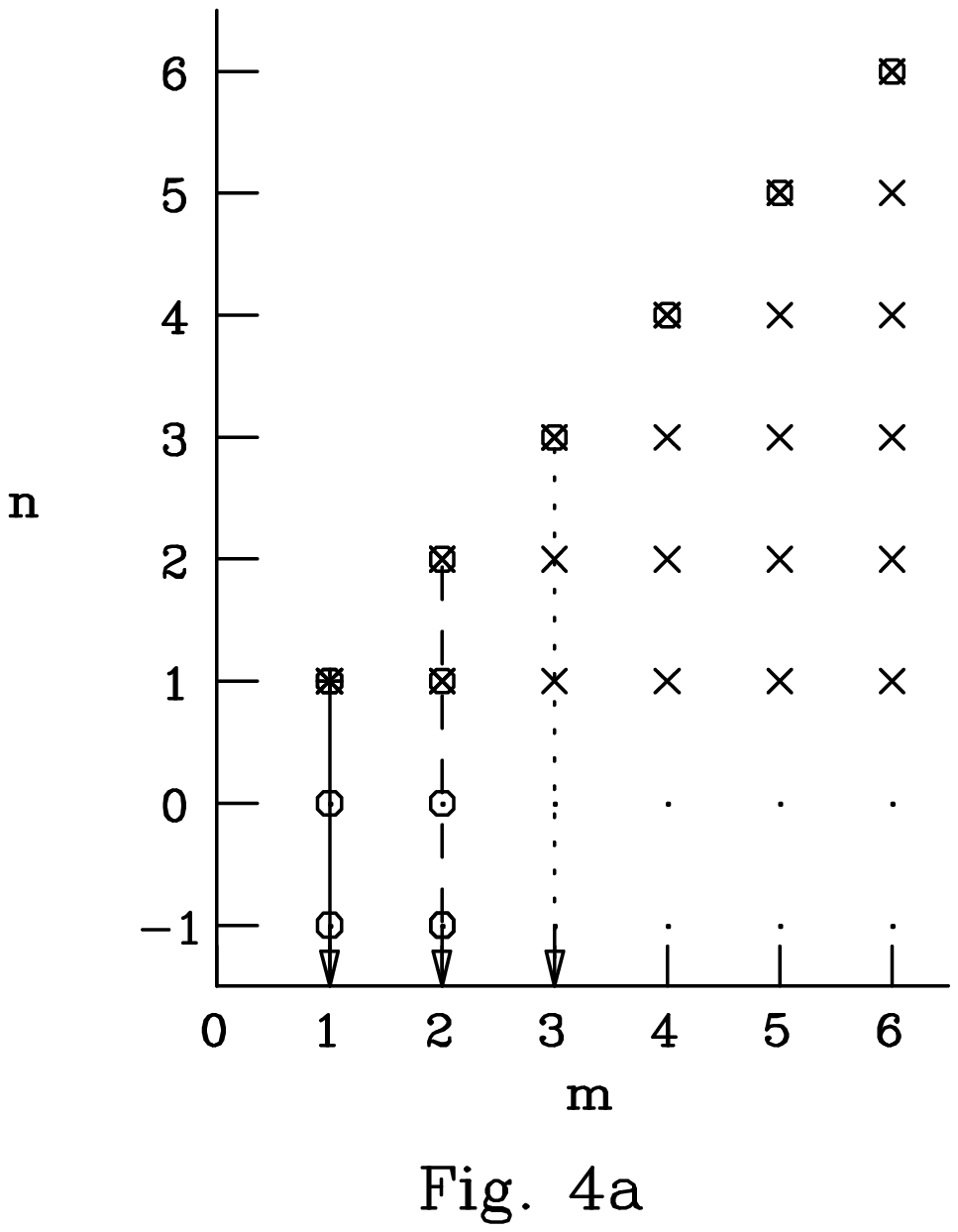}\hskip-5truecm
\hfil\epsfxsize=10truecm\epsfbox{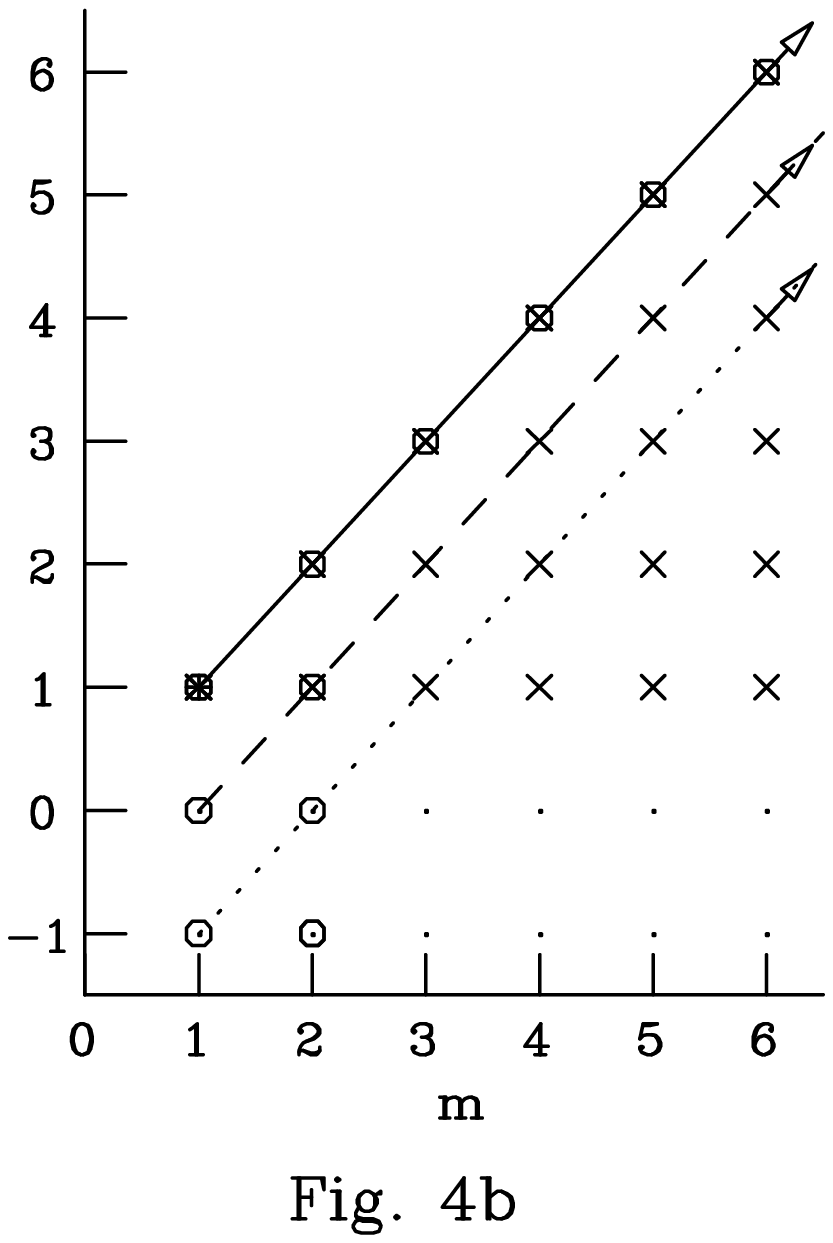}\hskip-5truecm
\hfil\epsfxsize=10truecm\epsfbox{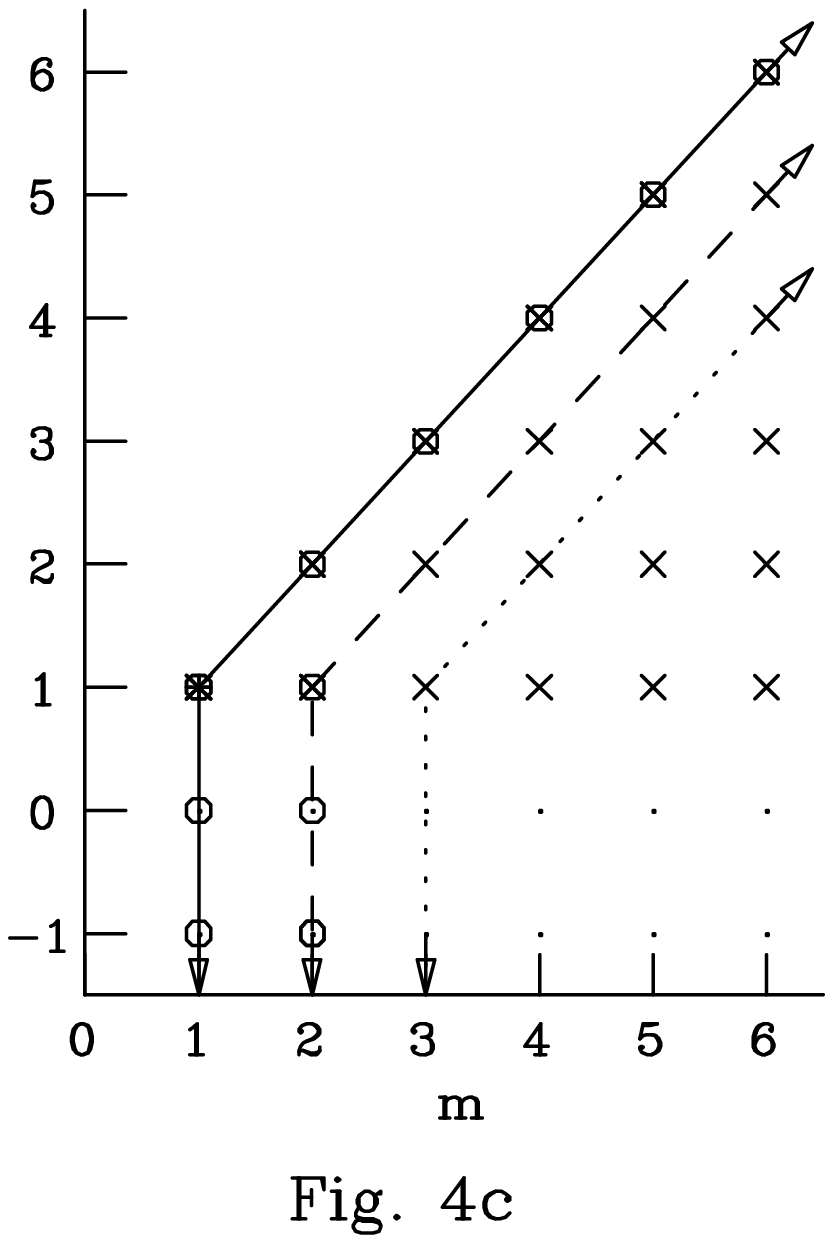}\hfil}
\vfill
\bye